\documentclass[journal]{IEEEtran}

\usepackage{cite}

\ifCLASSINFOpdf

\else

\usepackage[dvips]{graphicx}

\DeclareGraphicsExtensions{.eps} \fi

\usepackage[cmex10]{amsmath} 
\usepackage{amssymb}
\usepackage{flushend}

\usepackage{mdwmath} 
\usepackage{mdwtab}

\usepackage{eqparbox}

\usepackage{mathtools}

\usepackage[tight,footnotesize]{subfigure}

\allowdisplaybreaks
\DeclarePairedDelimiter{\ceil}{\lceil}{\rceil}

\newcommand{\abs}[1]{\left\lvert{#1}\right\rvert}

\newcommand{\norm}[1]{\left\lVert{#1}\right\rVert}

\newcommand{\inProd}[2]{\langle{#1},{#2}\rangle} 
\newcommand{\g}{\mid}

\newcommand{\mc}[1]{\mathcal{#1}} 
\newcommand{\mbb}[1]{\mathbb{#1}} 
\newcommand{\mbf}[1]{\mathbf{#1}} 
\newcommand{\st}[2]{\stackrel{#1}{#2}}

\newcommand{\pb}[1]{\left({#1}\right)}

\newcommand{\sqb}[1]{\left[{#1}\right]}

\newcommand{\cb}[1]{\left\{{#1}\right\}}

\newcommand{\nn}{\nonumber}

\newtheorem{theorem}{Theorem} 
\newtheorem{lemma}{Lemma} 
\newcommand{\Expec}[1]{\textbf{E}\left({#1}\right)} 
 
\begin{document}

%
\title{Recovery guarantees for multifrequency chirp waveforms in compressed
radar sensing} 

\author{ \IEEEauthorblockN{Nithin Sugavanam 
and Emre Ertin}\\ \IEEEauthorblockA{The Ohio State University}}

\maketitle
\begin{abstract}
Radar imaging systems transmit modulated wideband waveform to achieve high range
resolution resulting in high sampling  rates at the receiver proportional to the
bandwidth of the transmit waveform. Analog processing techniques can be used on
receive to reduce the number of measurements to $N$, the number of potential
delay bins. If the scene interrogated by the radar is assumed to be sparse
consisting of $K$ point targets, results from compressive sensing suggest that
number of measurements can be further reduced to scale with $K \log N$ for
stable recovery of a sparse scene from measurements with additive noise. While
unstructured random projectors  guarantee successful recovery under sparsity
constraints, they cannot be implemented in the radar hardware in practice.
Recently, structured random Toeplitz and Circulant matrices that result from
using stochastic waveforms in time delay estimation setting have been shown to
yield recovery guarantees similar to unstructured sensing matrices. However, the
corresponding transmitter and receiver structures have high complexity and large
storage requirements. In this paper, we propose an alternative low complexity
compressive wideband radar sensor which combines multitone signal chirp 
waveform on transmit with a receiver that utilizes an analog mixer followed with
a uniform  sub-Nyquist sampling stage.  We derive the recovery guarantees for
the resulting structured measurement matrix and  sufficient conditions for the
number of tones.  The only random component of our design is the sparse tone
spectrum implementable efficiently in hardware. Our analytical and empirical
results show that the performance of our scheme is in par with unstructured
random sensing matrices and structured Toeplitz and Circulant
matrices with random entries.
\end{abstract}

\begin{IEEEkeywords}
	Compressive sensing, mutual coherence, Structured measurement matrix,
Linear Frequency modulated waveform, Radar. 
\end{IEEEkeywords}

%
\IEEEpeerreviewmaketitle 
\section{Introduction} 
Radar imaging systems acquire information about the scene of interest by
transmitting pulsed waveforms and analyze the received backscatter energy to
form an estimate of the range and amplitude of the reflectors in the scene.
These range profiles from multiple pulses and/or multiple antenna elements can
be processed jointly to solve a multitude of inference tasks including
detection, tracking and classification~\cite{RadarSigProc}. In this paper, we
focus on the problem of estimation of range and amplitude of reflectors in the
scene using a single modulated wideband pulse $\phi(t)$ of bandwidth $B$. The
resolution of the echo imaging system is directly proportional  to the bandwidth
of the transmitted signal. Assuming the support of the observed delays are known
to lie on an interval $T$ (termed as range swath in radar literature), then the
unknown range profile can be discretized into $N=BT$ delay bins. The signal
model at the receiver can be  written as $y(t) =   \sum_{n=1}^N x_n \phi(t - n
\Delta)  + n(t)$, where $\Delta=1/B$, $n(t)$ is the receiver noise and  $x_n$
denotes the complex scattering coefficients associated with the $n'$th delay
bin.
Commonly the received signal is matched filtered with a copy of the transmitted
pulse to detect reflectors in range and estimate their complex amplitude of
backscattered energy. Direct digital implementation of the  matched filtering
step requires quadrature sampling of the received signal for the pulse duration
with sampling rate matching the bandwidth of the transmit waveform.
Alternatively, the matched filtering can be implemented in the analog domain
where the number of samples are reduced to $N$ to cover the delay support. 
However, the match filter output still requires to be sampled at the Nyquist
rate
corresponding to the system bandwidth, rendering digital and analog matched
filter receivers for {\em arbitrary} waveforms impractical for  bandwidths
exceeding gigahertz for current analog to digital converter (ADC) technology.
 
If a linear frequency modulated waveform (LFM) $\phi(t)=e^{j\beta t^2}$ is used 
on transmit, the matched filtering can be approximately implemented through
mixing the received signal with a reference LFM waveform and low pass filtering the mixer output. 
At the mixer output each copy of the waveform delayed by $\Delta$ appears as a tone whose 
frequency is given by  $\beta\Delta$. This pre-processing step is termed stretch processing
\cite{RadarSigProc, Middleton_DechirpMatchedFilter_2012} and can result in 
substantially reduced sampling rate for the ADC  used in the mixer output if the
delay support $T$ is smaller than the pulse length. 
Specifically, received signal  at stretch processor's output can be written as:
\begin{align}\label{eq:singleChirp}
 y(t) = \sum_{n=1}^N x_n e^{j (n\beta \Delta) t }, 
\end{align}
In essence, stretch processing converts range profile estimation problem into 
frequency spectrum estimation problem  with Nyquist rate  samples in time
obtained after analog processing.
If the scene can be assumed to have fewer targets $K$ than the number of delay
bins $N$, well-known results~\cite{DonohoCS_2006,RobustUncertainty_Candes_2006}
from the Compressive sensing (CS) shows successful reconstruction with 
sub-Nyquist samples is possible with the number of measurements $M$ scaling with
$K \log N$, if appropriate measurement operators can be implemented.  
Furthermore, there are numerous tractable algorithms with provable performance
that are either based on convex relaxation
\cite{Tao_2007_DantzigSel,ChenBP_2001,Tibshirani1996_LASSO} or greedy methods
\cite{SignalRecOMP_Tropp_2007,CoSaMP_Tropp_2009} to solve the reconstruction
problem.
Motivated by these advances, compressed sensing techniques have been  applied to
a variety of problems  in Radar~\cite{Potter2010}: range profile estimation
\cite{TimeDelay_unionSubspaces_Eldar_2010}, single
pulse systems for range-doppler estimation in \cite{HighResRadar_Strohmer_2009},
single pulse multiple transmit and receive system for range-doppler and azimuth
estimation and target detection in \cite{SparseMIMORadar,KerdockCodes}, remote
sensing in \cite{RemSense_Rauhut_2014}, direction of arrival estimation in
\cite{spatialMIMOCS_Eldar_2012}. CS based radar sensors  
based on pure random waveforms~\cite{NoiseRadar_Shastry_2015}, Xampling
framework~\cite{subNyquistRadar_Eldar_2014} and Random Modulator Pre-Integrator
(RMPI) framework~\cite{CSRadarHW_RMPI_2012}
using the receiver structure from \cite{RMPI_Tropp_2010} have also been
implemented. A common theme in most of the CS literature has been
randomization as it leads to measurement matrices that have provable recovery
guarantees. Implementation of the randomness into  compressive radar systems has
proven to be a challenging task in practice; uncorrelated random signals with
high peak-to-average power ratio is  mismatched to the nonlinear power
amplifiers used in radar systems and the system bandwidth (and as a result range
resolution) is limited as digital to analog converters (DAC) have to be employed
for generation of precise random signals for transmit and receive mixing. 

The LFM pulse model in \eqref{eq:singleChirp} provides an alternative strategy
as it converts the range estimation problem into an equivalent sparse frequency
spectrum estimation problem. Uniform subsampling in this setting has poor
perforamnce~\cite{SpectralCS_Duarte_2013}. Non-uniform random
sub-sampling can be used to obtain measurements with low mutual
coherency~\cite{FastFourierSampling_Tropp_2008,SpectralCS_Duarte_2013}. However,
 non-uniform sampling with commercially available ADCs still requires it to be
rated at the Nyquist rate to accommodate closely spaced samples.  We propose to
use low speed {\em uniform} sub-sampling using a high analog bandwidth ADC in
the sparse frequency spectrum estimation setting  and push randomness to
transmit signal structure to obtain compressive measurements. ADCs whose analog
bandwidth exceeds their maximum sampling rate by several
factors is readily available and used routinely in pass-band sampling.
This compressive radar structure proposed  in
\cite{FreqDivWaveform} uses linear combination of LFM waveform at the
transmitter with randomly selected center frequencies, while maintaining  the
simple standard stretch processing receiver structure. The output of the stretch
processor receiver is given by \[y(t) = \sum_{n=1}^{N}  x_n \sum_{k=1}^{N_c}
e^{j\phi_{n,k}}e^{ j \pb{n\beta \Delta + \omega_k} t }\]
where  $\phi_{n,k}$ is a predetermined known complex phase, $N_c$ is the number
of tones modulating the LFM waveform. We observe that under the proposed
compressive sensor  design each delayed copy of the transmitted waveform is
mapped to multi-tone spectra with known structure. As shown in this paper this
known multi-tone frequency structure enables recovery from aliased time samples
with provable guarantees. These results complement previous work which has 
shown good empirical performance in simulation and
measurements~\cite{sparseTargetMFMChirp}. 
\par
The contributions put forth in this sequel is that we present the theoretical
analysis for this multi-frequency LFM system. We show that the system with a
relatively small number of LFM waveform has performance guarantees similar to a
matrix with independent random entries for a sufficiently large number of tones
modulating the LFM waveform. We also present a numerical analysis comparing
our system with other measurement schemes. 
 \par
\subsection*{Notation and Preliminaries} We denote a vector in $N$-dimensional
complex domain as $\mbf{x} \in \mbb{C}^N$. $\norm{\mbf{x}}_0$ is called as
$\ell_0$ norm, which is given as the number of non-zero elements in a vector.
Clearly, this is not a valid norm but is used in formulating the fundamental
problem in compressed sensing. We denote $\norm {\mbf{x}}_1 = \sum_i \abs{x_i}$
as the $\ell_1$ norm. The Euclidean or $\ell_2$ norm is given as
$\norm{\mbf{x}}_2 = \sqrt{\sum_i \abs{x_i}^2 }$. We denote a matrix as $\mbf{A}
\in \mbb{C}^{M \times N}$, $\mbf{A}^{*} \in \mbb{C}^{N \times M}$ as conjugate
transpose of $\mbf{A}$, and $\mbf{I}$ as the identity matrix of dimensions
dependent on the context. The spectral or operator norm of the matrix is given
as $\norm{\mbf{A}}_{op} = \sigma_{\max}(\mbf{A})$ the largest singular value of
the matrix. The Frobenius norm of a matrix is given as $\norm{\mbf{A}}_F =
\sqrt{\sum_{i,j} \abs{A_{i,j}}^2}$. Another important quantity of interest is
the mutual coherence $\mu\pb{\mbf{A}}$, which is a measure of the correlation
between the columns of matrix $\mbf{A}$. The mutual coherence is given as
$\mu\pb{\mbf{A}} =\max_{i \neq j }
\frac{\abs{\inProd{\mbf{A}_i}{\mbf{A}_j}}}{\norm{\mbf{A}_i}_2
\norm{\mbf{A}_j}_2}$, where $\mbf{A_i}$ is a column of matrix $\mbf{A}$. Another
fundamental property
for the measurement matrices is called as Restricted Isometry Property (RIP). A
measurement matrix $\mbf{A} \in \mbb{C}^{M \times N}$ is said to satisfy RIP of
order $K$, if for any K-sparse vector $\mbf{x} \in \mbb{C}^N$ 
\begin{align}
	\label{eq:RIPCondition} &(1 - \delta) \norm{\mbf{x}}_2^2 \leq
\norm{\mbf{A}x}_2^2 \leq (1 + \delta) \norm{\mbf{x}}_2^2, \nn \\
	&\mbox{equivalently, } \delta_{K} =
\max_{\st{\mbf{\Gamma}}{card(\mbf{\Gamma}) \leq K}}
\norm{\mbf{A}_{\mbf{\Gamma}}^{*}\mbf{A}_{\mbf{\Gamma}} - \mbf{I} } \leq \delta, 
\end{align}
where $\mbf{\Gamma}$ is an index set that selects the columns of $\mbf{A}$, and
$card(\mbf{\Gamma})$ refers to the number of elements in the set, $\delta \leq
1$ and $\mbf{A}_{\mbf{\Gamma}}$ is the restriction of $\mbf{A}$ having columns
indexed by $\mbf{\Gamma}$. We denote the expectation operator as $\Expec{.}$.
The circularly-symmetric complex Gaussian distribution with mean $\mu$ and
variance $\sigma^2$ is denoted as $\mc{CN}\pb{\mu,\sigma^2}$.
\subsection{Relation to other works}
\begin{table*}[tbp]
\begin{center}
\begin{tabular}{| p{0.15\paperwidth} | c | c| c | p{0.12\paperwidth} | c |}
\hline \multicolumn{6}{|c|}{Recovery Guarantees from noisy measurements with
component-wise noise variance $\sigma^2$} \\
\hline Matrix Type of size $M \times N$  &  Mutual Coherence & Sparsity
for successful recovery & Spectral Norm & Minimum signal strength &
Reference
\\ \hline Random matrix with $(N M)$ independent random entries&
$2\sqrt{\frac{\log N}{M}}$ &
$\mc{O}\pb{\frac{M}{\log N}}$ &$\sqrt{\frac{N}{M}} + 1$ &
$\mc{O}\pb{\sigma\sqrt{2\log N }} $
&\cite{cai2011,candesPlanL1,GaussianOpNormConcentration}
\\ \hline Toeplitz block matrix with $(N+M)$ random entries 
&$\mc{O}\pb{\sqrt{\frac{\log N}{M}}}$ &$\mc{O}\pb{\frac{M}{\log N}}$
&$\mc{O}\pb{\sqrt{\frac{N}{M}}}$  &  $\mc{O}\pb{\sigma\sqrt{2\log N }} $ 
&\cite{Random_Teoplitz_Bajwa2012}
\\ \hline LFM waveform modulated with
 $N_c\ll N$ randomly selected tones& 
$\mc{O}\pb{\sqrt{\frac{\log N}{M}}}$
&$\mc{O}\pb{\frac{M}{\log N \log \pb{N + M}} }$  &$\mc{O}\pb{2 \sqrt{\frac{N
\log \pb{N + M}}{M}}}$ & $\mc{O}\pb{\sigma\sqrt{2\log N }} $ & This paper\\
\hline
\end{tabular}
\end{center}
\caption{Support recovery guarantees for different
sensing matrices.}
\label{table:SupportRec} 
\end{table*}
Related results  in literature on recovery guarantees for compressed radar sensing 
can be broadly categorized in two categories:  Results relating to  {\em signal reconstruction}
establish uniform recovery guarantees for  successful reconstruction
of all K-sparse signals, whereas  results on {\em support recovery} is concerned with the
detection of non-zero locations  of a  K-sparse signal that is assumed to
be sampled from a generic statistical model, such as uniformly sampling from all possible
 subsets of size $K$~\cite{candesPlanL1}.
 
In this paper  we  show that  randomly sampled  K-sparse signals  can be recovered with high probability using  LASSO  for  the structured  measurement matrix of the proposed 
ompressive radar sensor sensing scheme. Next, we  show that the
estimates of mutual coherence and column norms we obtain can be used to  provide 
uniform recovery guarantee following a standard argument.

Table~\ref{table:SupportRec} summarizes related results for support recovery
for different  measurement matrices. The upper bound on the sparsity level that
guarantees successful support recovery for our scheme has an
additional $\log(N + M)$ penalty compared to unstructured Gaussian matrix as
shown by \emph{Candes and Plan} in \cite{candesPlanL1} and block Toeplitz
matrices with entries sampled from Rademacher distribution as shown by
\emph{Bajwa} in \cite{Random_Teoplitz_Bajwa2012}.  We note that the random
matrix with independent entries is not realizable in radar setting but included in 
the table to provide a baseline.  \par

Uniform recovery guarantees are often  formulated in terms of  satisfying  RIP
property with high probability,
since if a measurement matrix satisfies RIP of order 2K such that $\delta_{2K}
\leq \delta \approx \sqrt{2} -1$,  then all K-sparse vectors are successfully
recovered with a reconstruction error of an oracle
estimator that knows the support of the sparse vector or the support of K
largest elements~\cite{RIP_compressedSensing_Candes2008}. 

\emph{Baraniuk et. al.} in \cite{RIPRandom_Baraniuk2008} have shown that random 
matrices with i.i.d entries from either Gaussian or sub-Gaussian probability
distribution satisfy 
the RIP condition. For any $\delta \in [0,1]$, $\delta_K \leq \delta$ if
$K \leq \alpha_4 \frac{M}{\log \pb{N/K}}$ where $\alpha_4$ is dependent on
$\delta$
and the sub-Gaussian norm of the random variables. Although these unstructured
random matrices have remarkable recovery guarantees they do not represent any
practical measurement scheme, which leads us to consider classical linear time
invariant (LTI) systems. \par Typically, an active imaging system transmits a
signal that interacts with a scene of interest and the acquired measurements are
used to estimate characteristics of the scene. The unknown
environment is modeled as an LTI system whose transfer function has to be
estimated using compressed measurements from the data acquisition step. It is
assumed that there exists a sparse or compressible representation of the
transfer function in some domain and the goal is to solve the sparse estimation
problem with the least possible measurements. This leads to a structured
measurement matrix that is either a partial or sub-sampled Toeplitz or circulant
matrix. The RIP condition of order $K$ for partial Toeplitz matrices in the
context of channel estimation was established by \emph{Haupt et. al.} in
\cite{Teoplitz_Haupt2010}. They showed that if the sparsity $K \leq \alpha_5
\sqrt{\frac{M}{\log N}}$, then $\delta_K \leq \delta $, where $\alpha_5$ depends
on $\delta$.
This quadratic scaling of number of measurements with respect to sparsity was
improved in \cite{RandomConvolution_Romberg2009,
RIP_toeplitz_Rauhut2012,Chaos_Rauhut2014}.
\emph{Romberg} in \cite{RandomConvolution_Romberg2009} considered an active
imaging system that used waveform with a random symmetric frequency spectrum and
acquired compressed measurements using random sub-sampler or random demodulator
at the receiver to estimate the sparse scene. The resultant system is a randomly
sub-sampled circulant matrix representing the convolution and compression
process. It is shown that for a given sparsity level $K$, the condition that
$\delta_{2K} \leq \delta$ is satisfied if the number of measurements $M \geq
\alpha_6 \delta^{-2} \min\pb{K (\log N)^6, (K \log N)^2 }$, where $\alpha_6 > 0$
is a universal constant independent of the size of problem and $\delta$. This
was extended  by \emph{Rauhut
et. al.} in \cite{RIP_toeplitz_Rauhut2012}. They consider a deterministically
sampled random waveform in time domain with samples following Rademacher
distribution, which is modeled as a sub-sampled Toeplitz or Circulant matrix
with entries sampled from Rademacher distribution. It was shown that for a given
sparsity level K, $\delta_K \leq \delta$ with high probability if the number of
measurements $M \geq \alpha_7 \max\pb{\delta^{-1}(K\log N)^{3/2},\delta^{-2} K
(\log N \log K)^2}$, where $\alpha_7$ is a universal constant. In the subsequent
work by \emph{Krahmer et. al.} in \cite{Chaos_Rauhut2014}, the relation between
sparsity level and number of measurements is improved and more general random
variables are considered such as vectors following sub-Gaussian distribution to
generate the Toeplitz or Circulant matrix. It is shown that, for a given
sparsity level K the condition $\delta_K \leq \delta$ is satisfied if the number
of measurements $M \geq \alpha_{8} \delta^{-2}K (\log K \log N)^2$, where the
constant $\alpha_{8}$ is a function of only the sub-Gaussian norm of the random
variables generating
the matrix. We adopt a method similar to \cite{Teoplitz_Haupt2010} and establish
the RIP condition of order K and obtain a similar result stating that $\delta_K
\leq \delta$ if $M \geq a \delta^{-2}K^2 log N$, where $a >0$ is independent of
$\delta$. \par

The rest of the paper is organized as follows,
Section~\ref{sec:Model} gives the mathematical model for the multi-frequency
chirp model and the statistical model considered for the target.
Section~\ref{sec:RecGuarantees} states the main result about the measurement
scheme employed for sparse recovery. Section~\ref{sec:Results}
contains detailed simulation results of the proposed multi-frequency chirp
waveform. Section~\ref{sec:Proofs} contains the detailed proof of the main
recovery result. We conclude with some future directions in
Section~\ref{sec:Conclusion}.

\section{Signal model and problem statement}\label{sec:Model} 
\subsection{Multi-frequency chirp model} We consider a radar sensor with 
collocated transmitter and receiver antennas employing the
compressive illumination framework proposed in \cite{FreqDivWaveform} and
\cite{sparseTargetMFMChirp} for
estimating the  range and complex reflectivity of reflectors in the scene.
The chirp rate of all the transmitted linear frequency modulated (LFM) waveform
is fixed at $\frac{\beta}{\tau}$, where $\beta$ is the bandwidth of each
transmitted waveform, $\tau$ is the pulse duration and $B = g \beta$ is the
system bandwidth for $g \geq 1$. We denote the unambiguous time interval as $t_u
= t_{\max}- t_{\min}$, where $t_{\max} = \frac{2R_{\max}}{c}$, $t_{\min} =
\frac{2R_{\min}}{c}$, $R_{\max}, R_{\min}$ are the maximum and minimum range in
the area of interest, respectively, while $c$ is the velocity of light in
vacuum. The whole space of range is discretized into grids based on the Radar's
resolution, therefore we get $N = B t_u$ grids. The interval of frequency from
$\sqb{0,B}$ is divided into $N$ grids such that $f(i) =\frac{iB}{N},
i=0,\cdots,N-1$, which are used as center frequencies for the chirp waveform.
From the possible $N$ waveform, a subset of size $N_c$ is chosen at random for
transmission. We simplify this selection model by considering independent
Bernoulli random variables as indicator variables to select LFM waveform such
that $N_c$ waveform are selected on an average. Let $\gamma_i \in\cb{0,1}$ be
the random variable indicating that $f(i)$ is part of the subset of size $N_c$.
It can be seen that 
\begin{align}
	\gamma_i = 
	\begin{cases}
		0 &\mbox{ with probability (w.p.) } 1 -\frac{N_c}{N} \\
		1 & \mbox{w.p. } \frac{N_c}{N} . 
	\end{cases}
\end{align}
The chosen LFM waveform are then scaled by independent random variables such as 
\begin{enumerate}
	\item a sequence of independent and identical complex phase with
probability density function as $f_{\Phi}(\phi_i) = \frac{1}{2\pi} ,\phi_i \in
\sqb{0,2\pi}$ , 
	\begin{align}
		\label{eq:BerUnifPhase} c_i = \gamma_i \exp(j\Phi_i), 
	\end{align}
	\item a sequence of scaling variables following Rademacher distribution
given by 
	\begin{align}
		\label{eq:BerRademacher} \xi_i &= 
		\begin{cases}
			-1 &\mbox{w.p. } 0.5 \\
			1 & \mbox{w.p. } 0.5\\
		\end{cases}
		\nn \\
		c_i &= \gamma_i \xi_i. 
	\end{align}
\end{enumerate}
We choose the model in \eqref{eq:BerRademacher}, which states that the $N_c$
chirp waveform are scaled by random signs in our analysis but use the model in
\eqref{eq:BerUnifPhase} in simulation results. \par The transmitted signal can
be written as 
\begin{align*}
	s(t) = \frac{1}{\sqrt{ M N_c}}\sum_{i=0}^{N-1} c_i \exp\pb{j2\pi (f_c +
\frac{i\beta}{M}) t + \frac{\beta}{2\tau}t^2 }, 
\end{align*}
where $0\leq t \leq \tau$. The receiver utilizes stretch processing
 at the same chirp rate as the transmitter and a fixed reference
frequency $f_d = f_c$ to demodulate the carrier frequency and estimate the
round-trip delay. The overall duration of the de-chirping waveform is $t_u
+\tau$. The sampling rate employed at the receiver is $F_s =
\frac{\beta}{\tau}t_u$. The total number of samples in the pulse duration $\tau$
is $M = \beta t_u$. The output samples of the stretch processor due to the
target at different delay bins $\Delta_m = \frac{m}{g \beta}$ are
\begin{align*}
	y(k) &= \frac{1}{\sqrt{M N_c}}\sum_{i=0}^{N-1} \sum_{m=0}^{N-1} c_i
\exp\pb{-j 2\pi \frac{im}{N} } \nn \\
	& \times \exp \pb{2\pi j \pb{\frac{i p}{M } - \frac{m}{N}} k } x(m) +
w(k), 
\end{align*}
where $k=0,\cdots,M-1$, $w_k$ is measurement noise process with 0 mean and
variance $\sigma^2$, $p = \frac{ \tau}{t_u} \in \mbb{Z}$ and $x\pb{m}$ is the
complex scattering coefficient due to a target at the delay bin $\Delta_m$. This
can be
compactly written as 
\begin{equation}
	\label{eq:singleTrModel} \mbf{y} = \mbf{Ax} + \mbf{w}, 
\end{equation}
where $\mbf{y},\mbf{w} \in \mbb{C}^{M}$, $\mbf{A} \in \mbb{C}^{M \times N}$, and
$\mbf{x} \in \mbb{C}^{N}$. The sensing matrix $\mbf{A}$ can be represented as a
series of deterministic matrices corresponding to the response to each of the
chirp waveform scaled by zero mean random coefficients as shown 
\begin{align}
	\label{eq:measurement1T_1R} \mbf{A} = \sum_{i=0}^{N-1} c_i \mbf{H}_i
\bar{\mbf{A}} \mbf{D}_i. 
\end{align}
The individual components are as follows 
\begin{align}
	\label{eq:individualChirpMatrices} &\bar{\mbf{A}} = \frac{1}{\sqrt{M
N_c}}
	\begin{bmatrix}
		\bar{\mbf{A}}(0) &\cdots &\bar{\mbf{A}}(N-1) 
	\end{bmatrix}
	\nn \\
	&\bar{\mbf{A}}(r) = 
	\begin{bmatrix}
		1 &\exp \pb{-2\pi j \frac{r}{N} } &\cdots &\exp \pb{-2\pi j
\frac{r(M-1}{N} ) } 
	\end{bmatrix}^T
	, \nn \\
	&\mbf{D}_i = 
	\text{diag}\begin{bmatrix}
		1 &\exp\pb{-j 2\pi \frac{i}{N}} &\cdots &\exp\pb{-j 2\pi
\frac{i(N-1)}{N}}
	\end{bmatrix}
	, \nn \\
	&\mbf{H}_i = 
	\text{diag}\begin{bmatrix}
		1 &\exp\pb{j 2\pi \frac{i p}{M } } &\cdots &\exp\pb{j 2\pi
\frac{i p(M-1)}{M }}
	\end{bmatrix}, 
\end{align}
where $i=0,\cdots,N-1$ and $r = 0,\cdots,N-1$, $\bar{\mbf{A}} \in \mbb{C}^{M
\times N}$ are the samples from tones that correspond to each delay bin
generated as a result of the de-chirping process in case of a single chirp
system, $\mbf{H}_i \in \mbb{C}^{M
\times M}$ is the shift in frequency due to the $i^{th}$ chirp waveform, and
$\mbf{D}_i \in \mbb{C}^{N \times N}$ is the phase term associated
with different delay bins due to the $i^{th}$ chirp. Each column of sensing
matrix $\mbf{A}$ can also be represented as 
\begin{align}
	\label{eq:individualColumnMatrices} &\mbf{A}(m) = \mbf{E}_m \mbf{F}
\mbf{G}_m \mbf{c}, \text{ where } \nn \\
	&\mbf{E}_m = 
	\text{diag}\begin{bmatrix}
		1 &\exp\pb{-j 2\pi \frac{m}{N} } &\cdots &\exp\pb{-j 2\pi
\frac{m (M-1)}{N}}
	\end{bmatrix}
	\nn \\
	&\mbf{F} = \frac{1}{\sqrt{M N_c}}
	\begin{bmatrix}
		\mbf{F}(0) &\cdots &\mbf{F}(N-1) 
	\end{bmatrix}
	\nn\\
	&\mbf{F}(r) = 
	\begin{bmatrix}
		1 &\exp \pb{2\pi j \frac{r p}{M} } &\cdots &\exp \pb{2\pi j
\frac{r p(M-1}{M} ) } 
	\end{bmatrix}^T
	  , \nn \\
	  &\mbf{G}_m = 
	  \text{diag}\begin{bmatrix}
		  1 &\exp\pb{-j 2\pi \frac{m}{N} } &\cdots &\exp\pb{-j 2\pi
  \frac{m (N-1)}{N}}
	  \end{bmatrix} 
\end{align}
where $m = 0,\cdots, N-1$, $r = 0,\cdots,N-1$, $\mbf{E}_m \in \mbb{C}^{M \times
M}$ represents the tone generated due to target present at $m^{th}$ delay bin,
$\mbf{F} \in \mbb{C}^{M \times N}$ are the different chirp center frequencies,
$\mbf{G}_m \in \mbb{C}^{M \times M}$ is the phase term due to different chirp
frequencies for a particular delay bin m, and $\mbf{c} \in \mbb{C}^N$ is the
random vector that selects the chirp waveform and scales them. A closer
inspection of matrix $\mbf{F}$ reveals that each of the center frequencies used
to shift the chirp waveform is being aliased into lower frequency tones as we
are sampling at Sub-Nyquist rate. We assumed $p \in \mbb{Z}$ in order to
simplify the analysis as we get sub-sampled Discrete Fourier Transform (DFT)
matrices. We impose an additional condition that $p$ should be co-prime with $M$
in order for $N$ frequency tones to be uniformly mapped
onto $M$ frequency bins, where $M \leq N$. A simple example of $p=M+1$, which
makes p co-prime with M, circularly maps the N possible frequencies into M
bins. 
\subsection{Target model} We consider a statistical model similar to \emph{
Strohmer and Friedlander} in \cite{SparseMIMORadar} for the sparse range profile
of targets. We assume that
the targets are located at the N discrete locations corresponding to different
delay bins. The support of the K-sparse range profile is chosen uniformly from
all possible subsets of size $K$. The complex amplitude of non-zero components
is assumed to have an arbitrary magnitude and uniformly distributed phase in
$\sqb{0,2\pi}$. 
\subsection{Problem statement} Given a sparse scene with targets following the
statistical model discussed in previous section, and measurement scheme in
\eqref{eq:singleTrModel} with $M \ll N$ and sparsity level $K \ll N$, the goal
of compressed sensing \cite{DonohoCS_2006} is to recover the sparse or
compressible vector $\mbf{x}$ using minimum number of measurements in $\mbf{y}$
constructed using random linear projections. The search for sparsest solution
can be formulated as an optimization problem given below 
\begin{align}
	\label{eq:l0_minimization} \min_{\mbf{x}} \norm{\mbf{x}}_0, \mbox{
subject to } \norm{\mbf{A}\mbf{x} - \mbf{y} }_2 \leq \eta, 
\end{align}
where $\eta^2$ is the noise variance. This was shown to be NP-hard and hence,
intractable
\cite{FoucartMathIntroCS_2013}, and many approximate solutions have been found.
One particular solution is to use convex relaxation technique to modify the
objective as $\ell_1$ norm minimization instead of the non-convex $\ell_0$ norm
given by, 
\begin{align}
	\label{eq:l1Relaxation1} \min_{\mbf{x}} \norm{\mbf{x}}_1 \mbox{ subject
to } \norm{\mbf{A}\mbf{x} - \mbf{y} }_2 \leq \eta. 
\end{align}
This approach has been shown to successfully recover sparse or compressible
vectors \cite{ChenBP_2001,RIP_compressedSensing_Candes2008} given that the
sub-matrices formed by columns of sensing matrix are well conditioned. Our
analysis is based on LASSO \cite{Tibshirani1996_LASSO}, which is a related
method that solves the optimization problem in \eqref{eq:l1Relaxation1}. It has
been shown in \cite{candesPlanL1} that for an appropriate choice of $\lambda$
and conditions on measurement matrix, the support of the solution of the below
mentioned
optimization problem coincides with the support of the solution of the
intractable problem in
\eqref{eq:l0_minimization}, 
\begin{align}
	\label{eq:l1Relaxation} \min_{\mbf{x}} \lambda \norm{\mbf{x}}_1 +
\frac{1}{2} \norm{\mbf{A}\mbf{x} -\mbf{y} }_2^2. 
\end{align}
The goal of our analysis is to show that the measurement model given
in \eqref{eq:measurement1T_1R} satisfies conditions on mutual coherence given
in\cite{candesPlanL1} and to find a bound on the sparsity level of range 
profile, which guarantees successful support recovery of almost all sparse
signals using LASSO with high probability from noisy measurements. In addition,
we also provide an estimate of the number of measurements required
for the sensing matrix representing our scheme to satisfy the RIP condition.
The next section presents our main results of our analysis.  
\section{Recovery guarantees}\label{sec:RecGuarantees} In order to obtain the
non-asymptotic recovery guarantee for our system employing multiple chirps, we
find an estimate of the tail bounds of mutual coherence and spectral or operator
norm of our measurement matrix. Using the estimates, we also provide conditions
for RIP condition of order $K$ to hold. \\
We make use of the Matrix Bernstein inequality given in
lemma~\ref{lem:MatBernstein} to get a tail bound on the operator norm for the
measurement matrix given in \eqref{eq:measurement1T_1R}. 
\begin{lemma}
	\label{lem:opNormTailBound} Given the measurement matrix model in
\eqref{eq:measurement1T_1R}, if $N_c \geq \frac{4}{9} \log(N + M)$, then we can
bound the tail probability for the operator norm as follows 
	\begin{align}
		\label{eq:MeasMatOpNorm} &P\pb{\norm{\mbf{A}}_{op} \geq
2 \sqrt{\frac{N \log \pb{N + M}}{M}}} \nn \\
		& \leq \pb{\frac{1}{N +M} }^{\alpha_2 -1}, \text{where }
	\end{align}
	\[\alpha_2 = \frac{2 }{1+\frac{2}{3}
\sqrt{\frac{\log \pb{N + M}}{N_c}}}.\]
	In addition, we also obtain an estimate of the expected value of the
operator norm of measurement matrix given as 
	\begin{align}
		&\Expec{\norm{\mbf{A}}_{op}} \leq \sqrt{\frac{2N}{M} \log\pb{N +
M} } + \frac{\log\pb{N+M}}{3}\sqrt{\frac{N}{M N_c}}. 
	\end{align}
\end{lemma}
The following results about the Euclidean norm of columns and mutual coherence
are obtained using concentration inequalities of quadratic forms of sub-Gaussian
random vectors given in \cite{HansonWrightSubGaussian}, which is extended to the
complex domain in lemma~\ref{lem:HansonWrightCom}. 
\begin{lemma}
	\label{lem:minColumnNormTailBound} The concentration inequality for the
minimum of Euclidean norm of any column $m$ of $\mbf{A}$ is given as follows 
	\begin{align}
		\label{eq:normColumnTail} &P\pb{\min_m \norm{\mbf{A}(m)}_2^2
\leq {1 - \epsilon}} \leq
\nn \\
		&4N
\exp\pb{-M d \pb{\epsilon
\frac{q^{*}}{\pb{\frac{N_c}{N}}^{\frac{2}{q^{*}} -1 }} }^2  } , 
	\end{align}
	where $d > 0$ is a universal constant, $\epsilon \in (0,1)$, and
	\[q^{*} = \max\pb{1,2\log\pb{\frac{N}{N_c}}}.\]
\end{lemma}
\begin{lemma}
	\label{lem:mutualCoherenceTailBound} If $M \geq (\log N)^3$ then there
exists constant $\alpha_3 > 0$ such that the mutual coherence of our sensing
matrix is bounded by 
	\begin{align}
		\label{eq:mutualCohTail} &P\pb{\mu\pb{\mbf{A}} \geq
\frac{\alpha_3}{1 - \epsilon} \sqrt{\frac{\log N}{M}} } \leq \nn
\\
		&
		\begin{cases}
			\frac{2}{N^{u_1-2} } + 4N \exp\pb{-M d\bar{\epsilon}^2 
}, & \mbox{if } \log N > \frac{q^{*}}{(\frac{N_c}{N})^{(2/q^{*}-1)}}
\alpha_3\\
			\frac{2}{N^{u_2-2} } + 4N \exp\pb{-M d \bar{\epsilon}^2
}, & \mbox{otherwise } 
		\end{cases}
	\end{align}
	where $d >0$ is a universal constant, $\epsilon \in
(0,1)$ are arbitrary
constants, and 
	\begin{align*}
		&u_1 = d \pb{\frac{q^{*}\alpha_3
}{\frac{N_c}{N}^{\pb{\frac{2}{q^{*}}-1}}}}^2,\\
		&u_2 = \frac{q^{*}\alpha_3
}{\frac{N_c}{N}^{\pb{\frac{2}{q^{*}}-1}}}d \log N,\\
		&q^{*} = \max\pb{1,2\log\pb{\frac{N}{N_c}}}, \\
		&\bar{\epsilon} =  \pb{\epsilon
\frac{q^{*}}{\pb{\frac{N_c}{N}}^{\frac{2}{q^{*}} -1 }} }. 
	\end{align*}
\end{lemma}
\begin{theorem}
	\label{thm:MultiChirpsRecGuarantee} For a measurement model $\mbf{y}=
\mbf{A}\mbf{x} +\mbf{w}$, where $\mbf{A}$ is defined in
\eqref{eq:measurement1T_1R} such that $\mbf{x}$ is drawn from a K-sparse model
in complex domain and $\mbf{w} \sim \mc{CN}(0,\sigma^2I)$, the following
conditions guarantee successful support recovery from solving
\eqref{eq:l1Relaxation} with regularizer $\lambda = 2\sigma \sqrt{\log N}$, 
	\begin{align}
		\label{eq:sparsityCondition} &K \leq K_{\max} =
\frac{\pb{1-\epsilon_1} \alpha_1
M}{  \log\pb{N} \log\pb{N + M}} , \\
		\label{eq:numMeasurementCondition} & M \geq \pb{\log N}^3, \log
N \geq \frac{q^{*}}{(\frac{N_c}{N})^{(2/q^{*}-1)}} \alpha_3 \\
		\label{eq:numChirpsCondition} & N_c \geq \max\pb{\frac{4}{9}
\log(N + M),\nu N }, \\
		\label{eq:minSignalCondition} & \min_{k \in I} \abs{x_k} >
\frac{8}{\sqrt{1-\epsilon}}\sigma\sqrt{2 \log\pb{N}}, 
	\end{align}
	with probability $\bar{p}_4( 1- \bar{p}_1 -\bar{p}_2 -\bar{p}_3)$
	%
	 for some $\alpha_3 > 0, \alpha_1 > 0, \epsilon \in
(0,1), \nu \ll 1$, $d > 0$ is a universal constant independent of N,M, where 
	\begin{align*}
		&\bar{p}_1 = \frac{2}{N^{u_1-2 }} + 4N
\exp\pb{-d M \bar{\epsilon}^2 } \\
		&\bar{p}_2 = \pb{\frac{1}{N +M} }^{\alpha_2 -1}+ 4N
\exp\pb{-d M\bar{\epsilon}^2 }\\
		&\bar{p}_3 = 4N \exp\pb{-dM \bar{\epsilon}^2 }, \\
		&\bar{p}_4 = 1- 2N^{-1}(2\pi \log \pb{N} + KN^{-1}) -
\mathcal{O} \pb{N^{-2\log 2}},\\
		&\alpha_2 = \frac{2 }{1+\frac{2}{3}
\sqrt{\frac{\log \pb{N + M}}{N_c}}}\\
		&u_1 = d \pb{\frac{q^{*}\alpha_3
}{\frac{N_c}{N}^{\pb{\frac{2}{q^{*}}-1}}}}^2\\
		&q^{*} = \max\pb{1,2\log\pb{\frac{N}{N_c}}},\\
		&\bar{\epsilon} =  \pb{\epsilon
\frac{q^{*}}{\pb{\frac{N_c}{N}}^{\frac{2}{q^{*}} -1 }} }. 
	\end{align*}
\end{theorem}
The proof in section~\ref{sec:Proofs} involves direct application of
lemma~\ref{lem:recoveryGuaranteePlan} and uses the estimates of the spectral
norm and the mutual coherence of the measurement matrix. 
\begin{theorem}\label{thm:RIPCondition}
 For the measurement matrix $\mbf{A}$ given in \eqref{eq:measurement1T_1R} and
 any $\delta, \epsilon \in \sqb{0,1}$ such that $\delta + \epsilon < 1$, the
RIP condition given as $\delta_K \leq \delta + \epsilon$ is satisfied with
probability $1- p_5 -p_6$ if the number of measurements $M \geq a \delta^{-2}
K^2 \log N$, where   
\begin{align*}
 &p_5 = \frac{1}{N^{\pb{u_3 - 2}}}, \\
 &p_6 =  4 N \exp\pb{-d
\pb{\epsilon\frac{q^{*}
}{\frac{N_c}{N}^{\pb{\frac{2}{q^{*}}-1}}}}^2 M } , \\
 &u_3 = a \pb{\frac{q^{*}
}{\frac{N_c}{N}^{\pb{\frac{2}{q^{*}}-1}}}}^2,
\end{align*}
$a >0$ is a constant independent of $N,M$. 
\end{theorem}
 We adopt a similar approach as \emph{Haupt et. al.} in
\cite{Teoplitz_Haupt2010} and utilize the estimates of inner-product of
columns of sensing matrix and norms to obtain a simple bound on the number of
measurements required to guarantee RIP of order K.

\subsection*{Discussion} The support recovery guarantee stated in
Theorem~\ref{thm:MultiChirpsRecGuarantee} is satisfied for almost all K-sparse 
vectors sampled from the generic sparse signal model discussed earlier, {\em i.e.} 
given a measurement matrix one could find a K-sparse vector (with arbitrarily small 
probability varying $N,M$ and $N_c$. )  for which the recovery fails.  
This differs from the worst-case guarantees  as well as reconstruction error bounds that depend on
Restricted Isometry Property (RIP) given in Theorem~\ref{thm:RIPCondition}. 
The exponent in probability tail bounds for quantities such as mutual
coherence $\mu(\mbf{A})$, spectral norm $\norm{\mbf{A}}_{op}$ of the measurement
matrix are controlled by number of chirp waveform employed
$N_c$. Specifically, it can be seen that the upper bound to the tail probability
of the above estimates reduce as $N_c$ increases until $N_c \leq \frac{N}{e}$.
We also show empirically in section~\ref{sec:Results} that the expected value of mutual
coherence $\mu(\mbf{A})$ reduces as the number of chirp waveform
increases. Specifically, it converges in mean to the mutual coherence of an
unstructured random matrix $\mbf{G}$ with independent Gaussian entries and 
thereby  converges in probability as well. Typically, smaller values of $\mu(\mbf{A})$ are desirable
for robust recovery as shown by \emph{Candes and Plan} in \cite{candesPlanL1} as
it ensures that the Grammian matrices of the sub-matrices formed using a subset
of columns of sensing matrix $\mbf{A}$ are well conditioned as shown by
\emph{Tropp} in \cite{normSubmatrices}. Since, the minimum value of the signal
has to be above the noise floor \eqref{eq:minSignalCondition} for successful
recovery, we get a condition on the signal to noise ratio SNR for a particular
target located at a fixed range bin $r$ below which the recovery guarantee does
not hold, which is given by 
\begin{align*}
	SNR_r = \frac{\abs{x_r}^2}{\sigma^2} \geq 128 \kappa\log N. 
\end{align*}
The authors in \cite{SparseMIMORadar,candesPlanL1} also show that the threshold
on $SNR_r$ scales with $C \log N$ with constant $C$
determining the probability of successful recovery.In section~\ref{sec:Results} we study the effect of SNR on the reconstruction error using simulations.
 
\section{Simulation examples} \label{sec:Results} For our simulation studies, we consider a 
system with a bandwidth $B=1GHz$ from which we choose center frequency of each chirp waveform
randomly with each chirp sweeping a fraction of the system bandwidth $\beta= \frac{1}{g} B$.
We note that a wideband  multi-tone signal  with bandwidth  $B$ modulating a LFM waveform of bandwidth $\beta$  results in a system bandwidth of $B+\beta$ Hz,  but we seek to 
resolve targets with range resolution that corresponds to  $B$ Hz common to all modulated chirps. 
The fractional bandwidth defined as the ratio of $\frac{\beta}{B}$ represents the under-sampling ratio as the 
stretch processor output  is uniformly sub-sampled at that rate. We assume that the minumum and
minimum range of the area of interest are $R_{\min}=0 m$ and $R_{\max} =150 m$ ,
respectively. The pulse duration $\tau$ is chosen such that the ratio $p =
\frac{\tau}{t_u} \in \mbb{Z}$ is co-prime with the number of samples. We make
use of the model in \eqref{eq:BerUnifPhase} to
select a subset of $N_c$ chirp waveforms and scale with a random phase term to
obtain the simulation results. The other measurement matrices that we compare
the performance of our scheme with are 
\begin{enumerate}
	\item matrix $\mbf{G}$ with i.i.d. complex Gaussian random entries
sampled from distribution $\mc{CN}(0,\frac{1}{M})$ , 
	\item matrix $\mbf{T} = \frac{1}{\sqrt{M}}\mbf{P}_{\Omega} \mbf{T}_1$,
which is a partial Toeplitz matrix, where $\mbf{P}_{\Omega}$ is a uniform
sub-sampling operator, 
	\begin{align*}
		&\mbf{T}_1 = 
		\begin{bmatrix}
			t_N & t_{N-1} & \cdots & t_1 \\
			t_{N+1} & t_{N} & \cdots & t_2 \\
			& \vdots & \vdots & \\
			t_{2N - 1} &t_{2N - 2} &\cdots &t_{N} 
		\end{bmatrix}
		\\
		& t_i \sim \mc{CN}(0,1), i=1,\cdots, 2N-1. \text{ and} 
	\end{align*}
\end{enumerate}
\par 
\begin{figure}
	[!t]\begin{center}
	\includegraphics[width=\linewidth]
{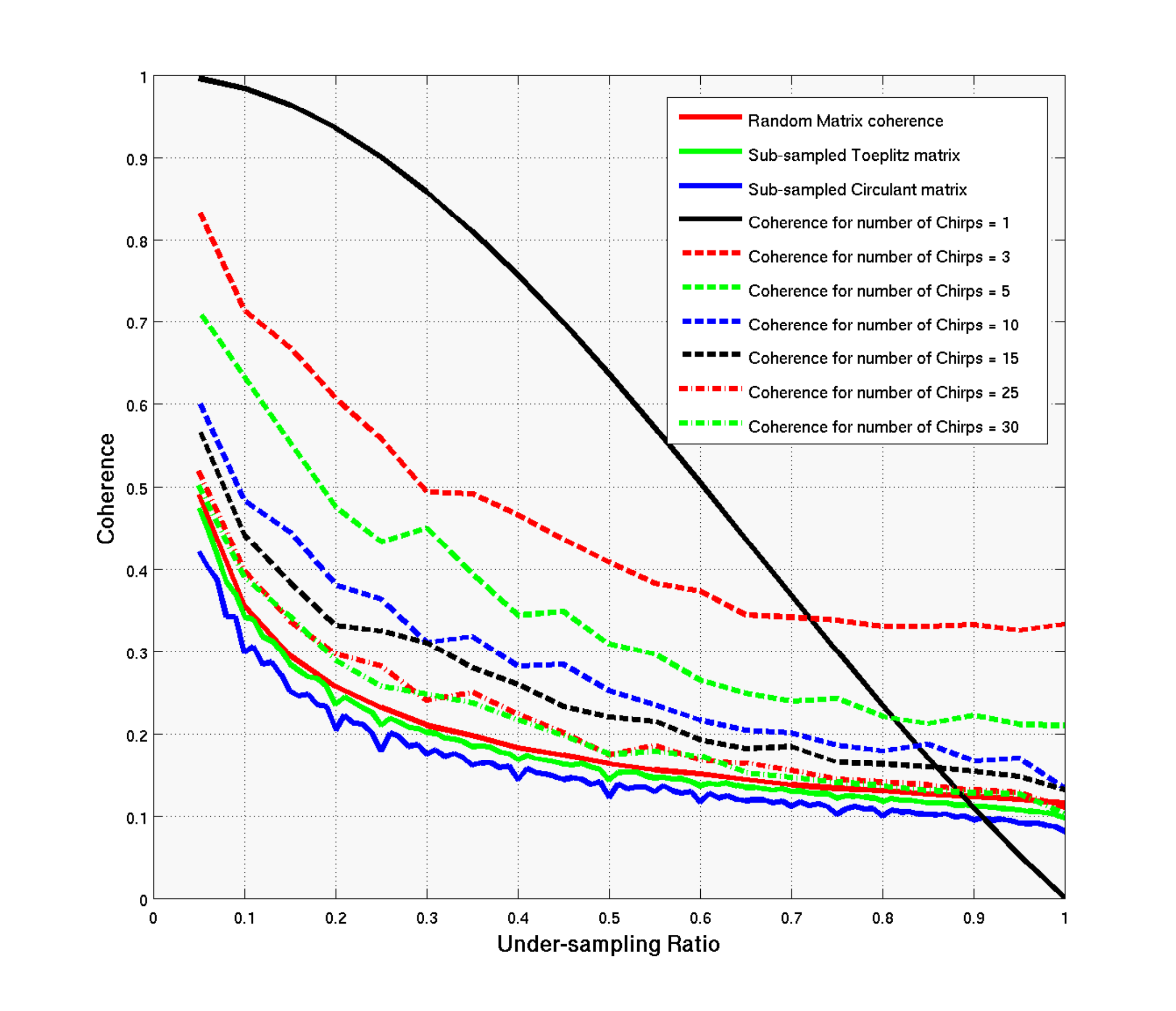} \caption{ Mutual Coherence of 
structured sensing matrices as a function of under sampling ratio. Random Gaussian matrix coherence is
provided as a baseline.~\label{fig:CohGraph} } 
\end{center}
\end{figure}
We generate 100 realizations of matrices $\mbf{G}, \mbf{A}$ and $\mbf{T}$, in
order to observe the effect of under-sampling on mutual coherence.
It can be seen empirically from figure~\ref{fig:CohGraph} that the mutual
coherence of the sensing matrix $\mbf{A}$ representing our system converges in
mean to the mutual coherence of a sensing matrix $\mbf{G}$ as the number of
chirps increase. We observe that even for small values of $\nu \st{\Delta}{=}
\frac{N_c}{N} =0.03$, the mean of the mutual coherence of our measurement matrix
$\mbf{A}$ is quite close to the mean of the mutual coherence of matrix
$\mbf{G}$. In
addition, we also evaluate the coherence of Toeplitz matrix $\mbf{T}$, which is
representative of active-imaging schemes using stochastic waveform that
are modeled as linear time invariant systems. \par 
\begin{figure*}
	[!t]
 \subfigure[$\mbf{A}$ with  $N_c=5$ Chirps]{
	\includegraphics[trim=0 0 0
0,clip,width=0.23\linewidth]{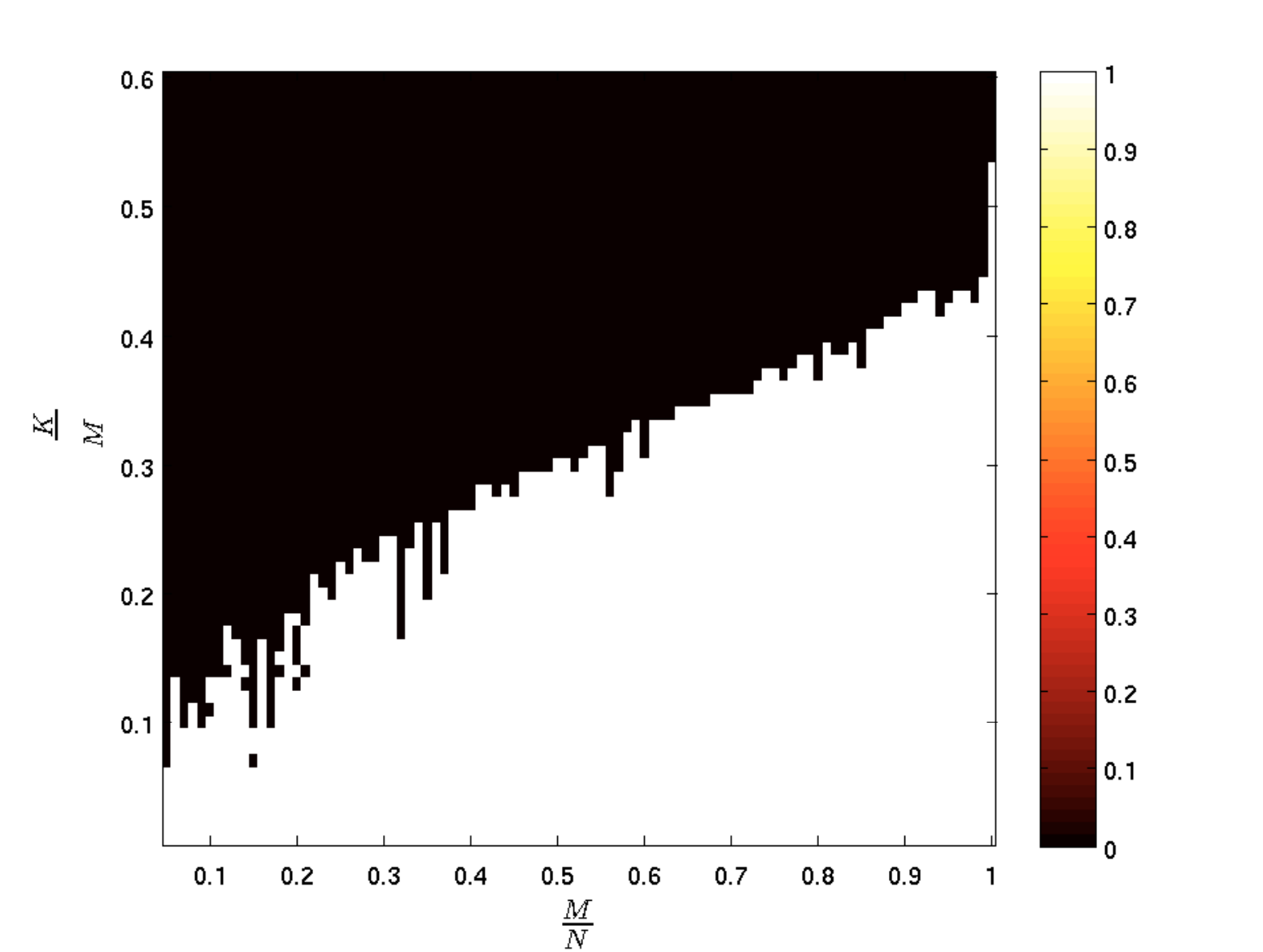} \label{fig:phaseTr_1t_1r_5}} 
	\subfigure[$\mbf{A}$ with  $N_c=25$ Chirps]{
	\includegraphics[trim=0 0 0
0,clip,width=0.23\linewidth]{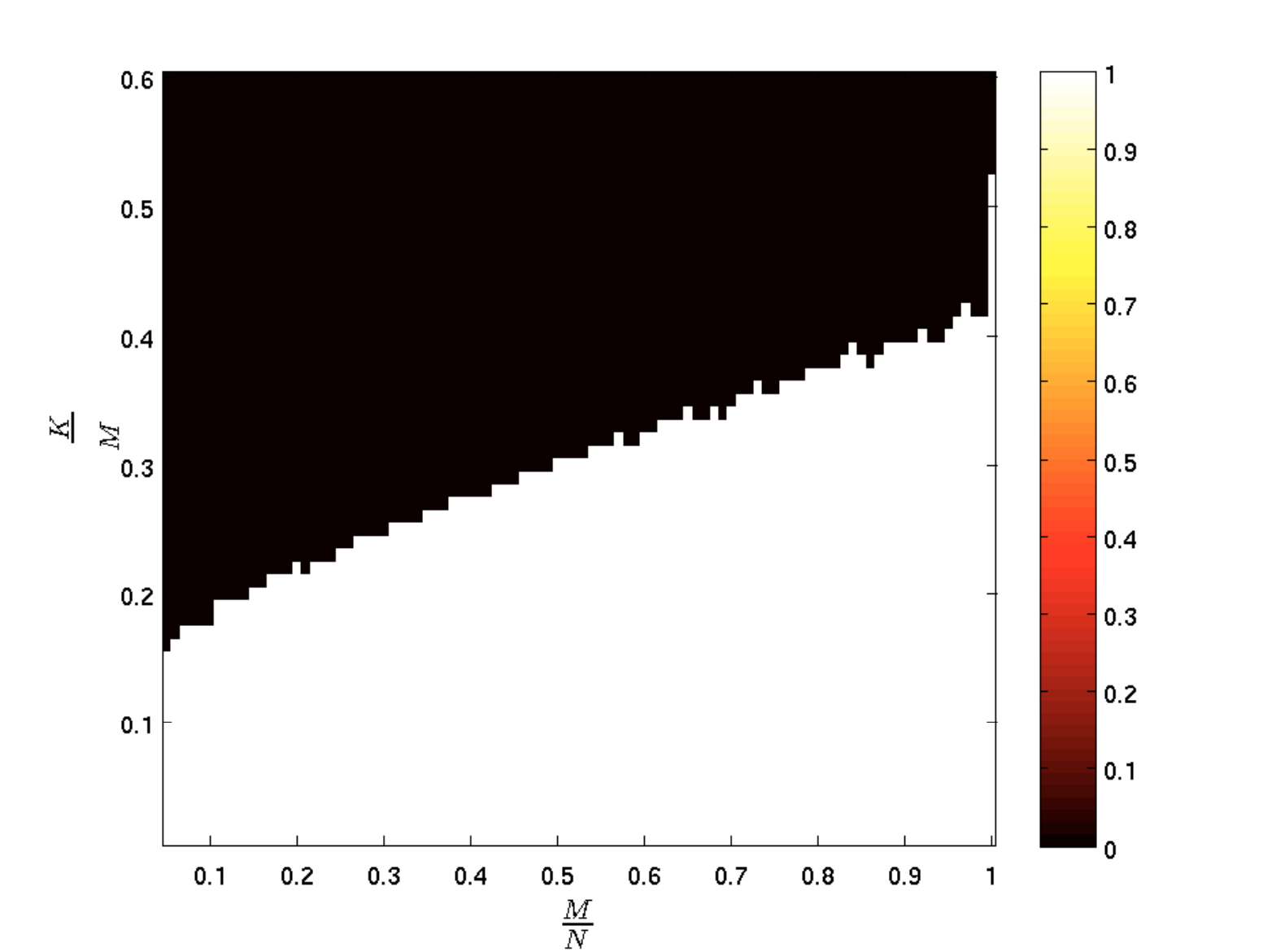} \label{fig:phaseTr_1t_1r_25}}
 \subfigure[Gaussian sensing matrix $\mbf{G}$.]{
	\includegraphics[trim=0 0 0
0,clip,width=0.23\linewidth]{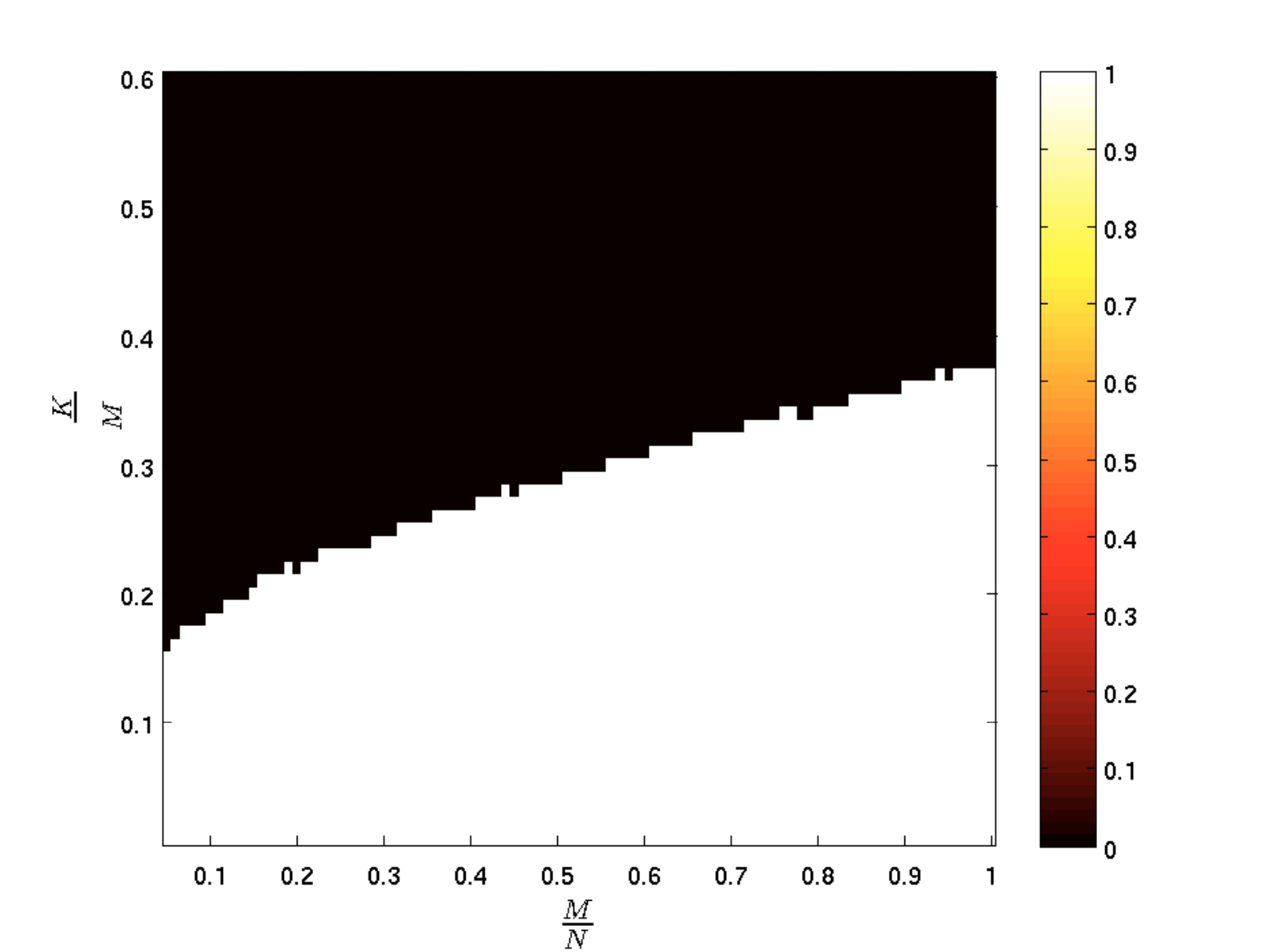}
\label{fig:phaseTr_Random}} 
	\subfigure[Toeplitz sensing matrix $\mbf{T}$.]{
	\includegraphics[trim=0 0 0
0,clip,width=0.23\linewidth]{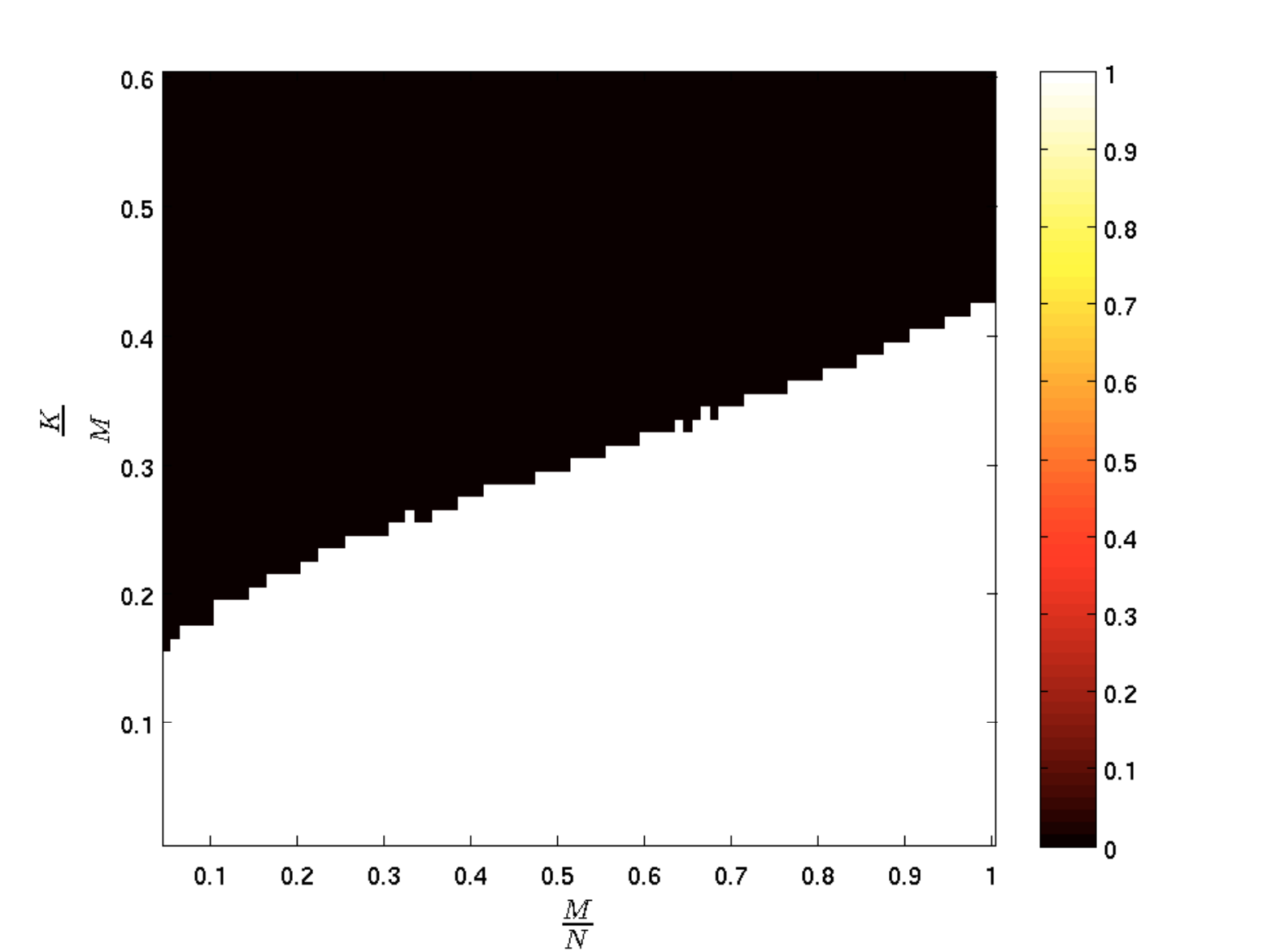} \label{fig:phaseTr_Toeplitz}}
	\caption{Phase transition curve for the multi-frequency measurement
scheme as number of chirps $N_c$ increases and other related sensing schemes.
Image intensity is either $1$ if reconstruction error is below
$1\%$ and $0$ otherwise. We compare the performance of our multi-frequency
scheme with other random structured and unstructured matrices as the sparsity
level and fractional bandwidth is varied at Signal to Noise ratio of 25 dB. }
\label{fig:RecoveryGuarantee}
\end{figure*}
Next, we consider the recovery performance of our measurement system used in
conjunction with Basis pursuit de-noise using SPGL1 solver developed by
\emph{Van den Berg and Friedlander} in \cite{spgSolver,spgSolver1} to estimate
the unknown target locations and their amplitudes in the area of interest. For
each realization of
the measurement matrix we generate multiple samples of target range profile with
specified sparsity level and scattering coefficient is sampled at specified
locations are sampled from a complex Gaussian distribution. The overall target
range profile is normalized to get a fixed SNR. We consider a function of mean
square error as a performance measure, specifically a thresholding function for
a fixed SNR of 25dB. We consider a threshold of $1\%$ on the mean squared error
and vary both the number of targets in the scene, and the bandwidth of the chirp
waveform, which in turn influence the sampling rate at receiver. Again it is
clear from figure~\ref{fig:RecoveryGuarantee} that the performance becomes
similar to that of the random Gaussian sensing matrix $\mbf{G}$ as the number of
chirps increase. We also observe that the recovery performance of Toeplitz
matrix $\mbf{T}$ is similar to our system at
lower values of fractional bandwidth $\frac{\beta}{B} \leq 0.7$ but performs
better when $\frac{\beta}{B} \geq 0.7$. \par 
\begin{figure*}	
	[!t]
 \subfigure[$\mbf{A}$ with  $N_c=5$ Chirps]{
	\includegraphics[trim=0 0 0
0,clip,width=0.23\linewidth]{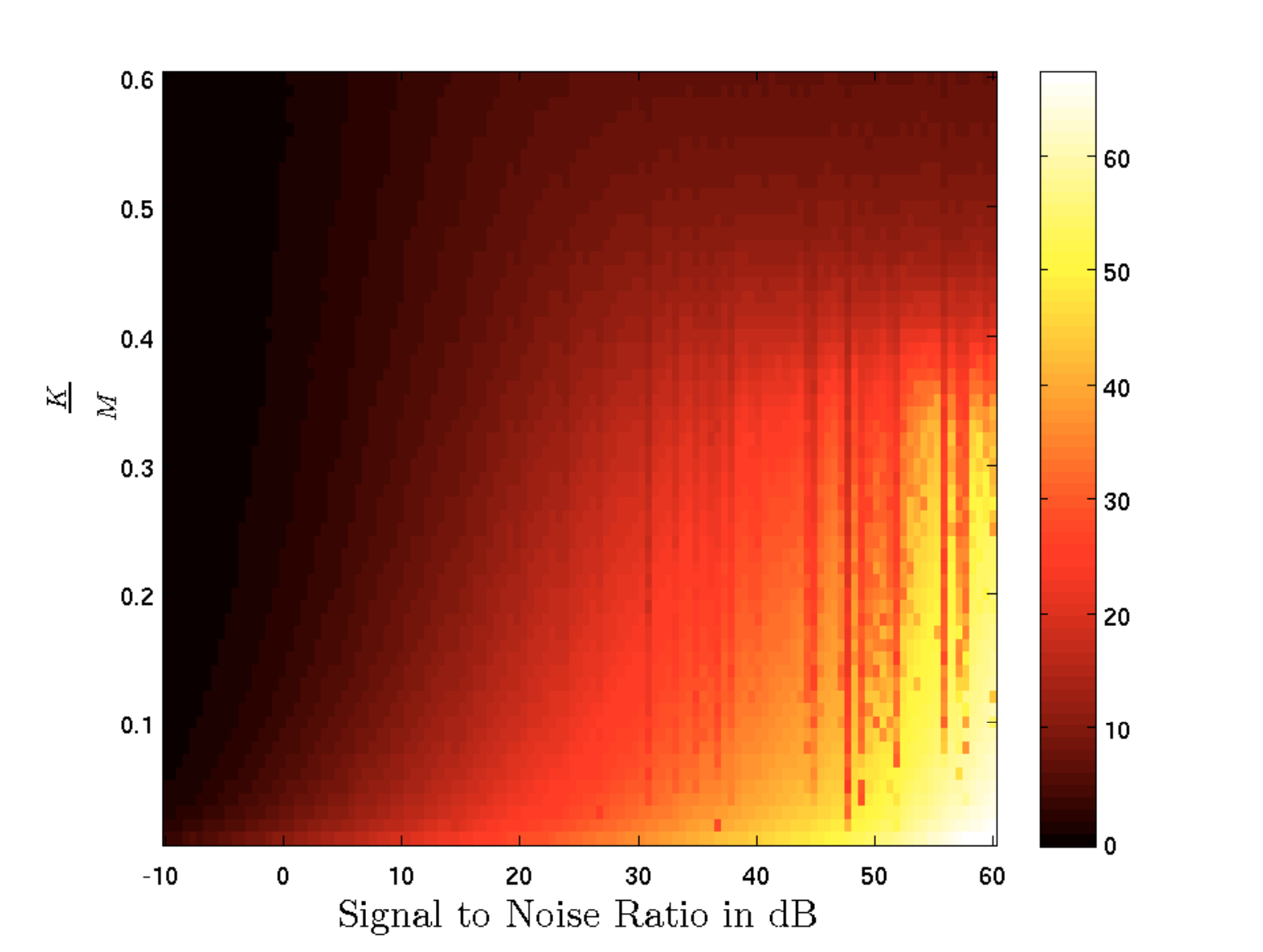} \label{fig:SNRvar_1t_1r_5}}
	\subfigure[$\mbf{A}$ with  $N_c=25$ Chirps]{
	\includegraphics[trim=0 0 0
0,clip,width=0.23\linewidth]{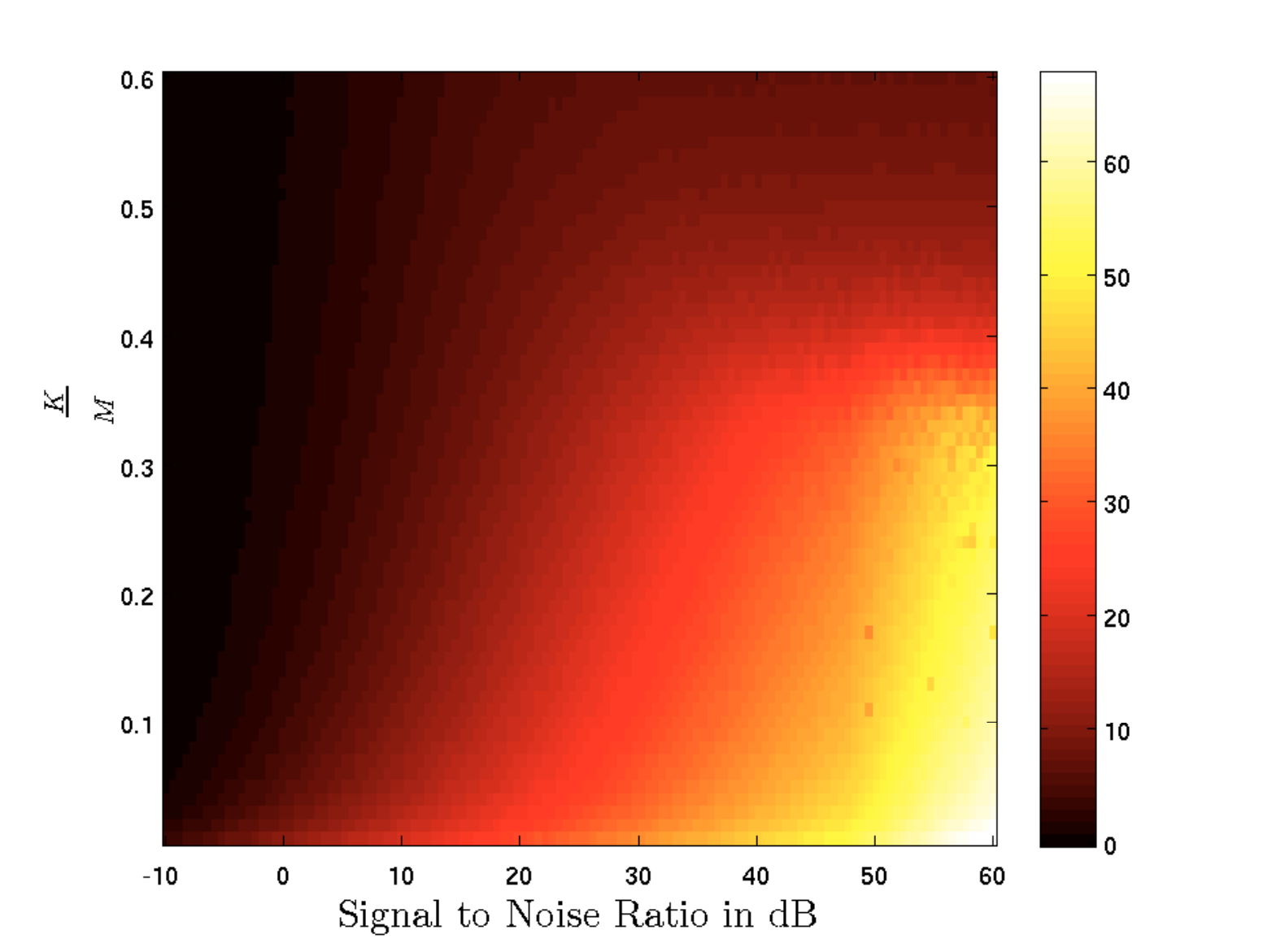}
\label{fig:SNRvar_1t_1r_15}}  \subfigure[Gaussian
sensing matrix $\mbf{G}$ ]{
	\includegraphics[trim=0 0 0
0,clip,width=0.23\linewidth]{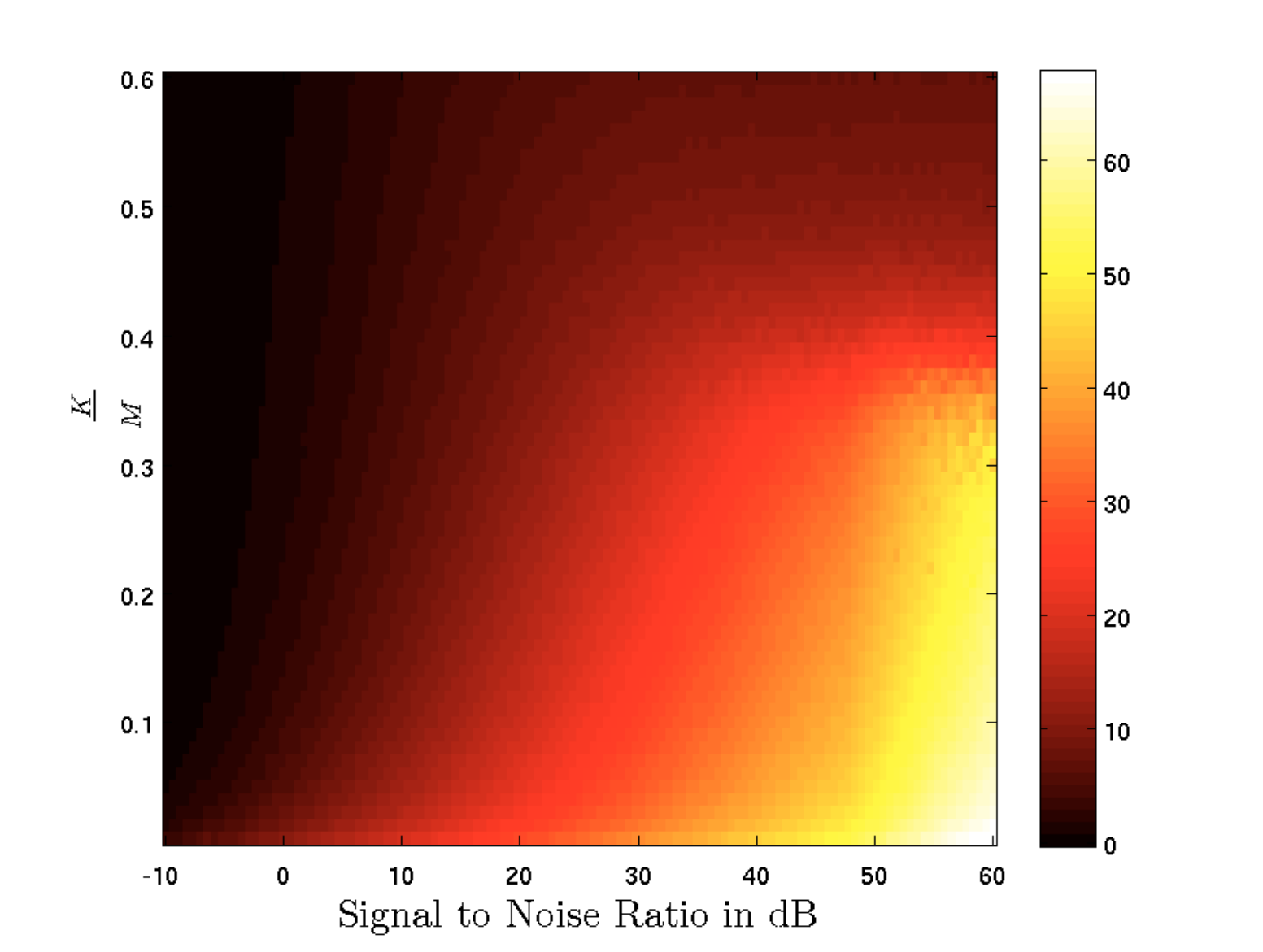} \label{fig:SNRvar_Random}}
	\subfigure[Teoplitz sensing matrix $\mbf{T}$ ]{
	\includegraphics[trim=0 0 0
0,clip,width=0.23\linewidth]{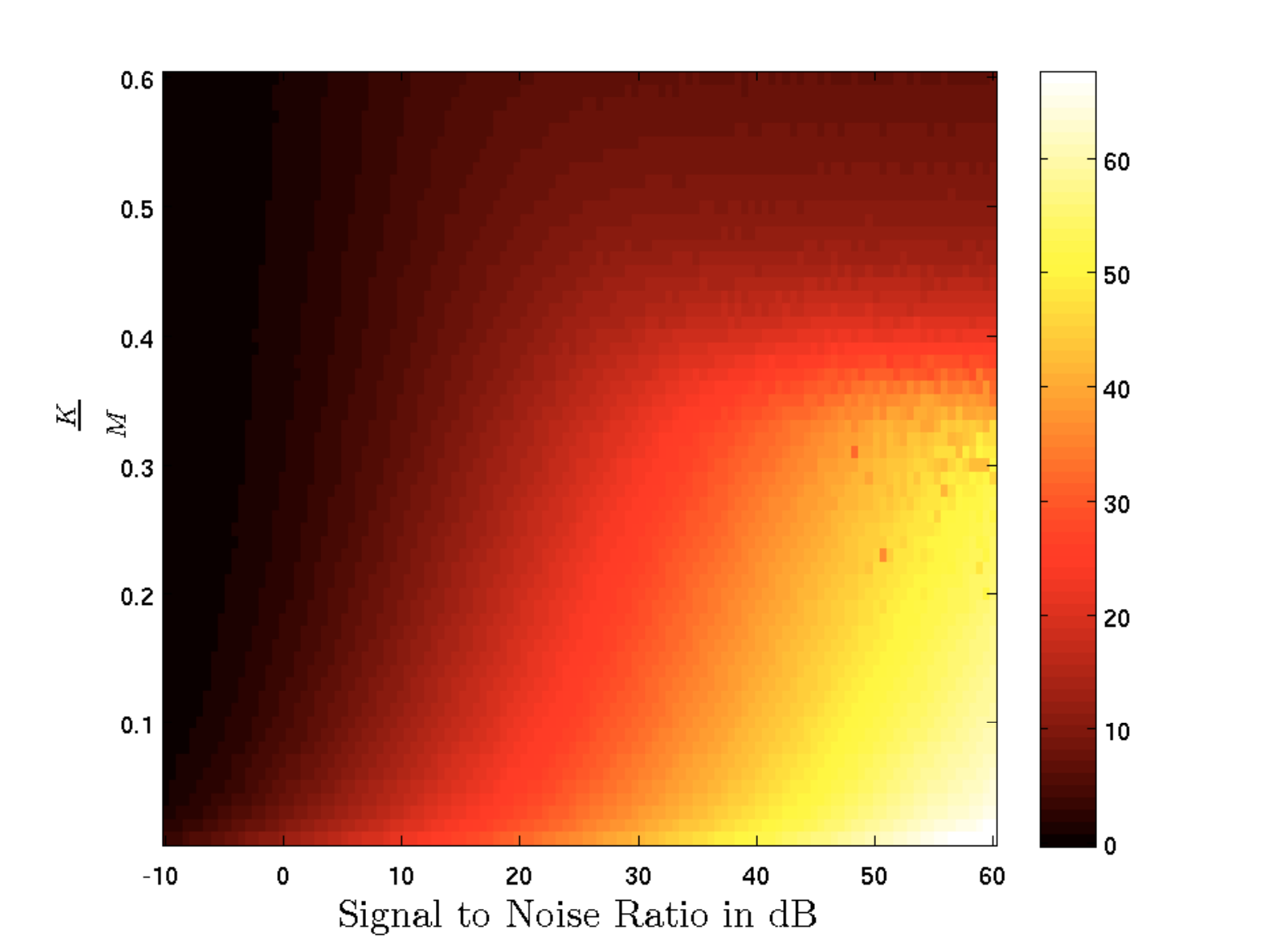} \label{fig:SNRvar_Toeplitz}}
	\caption{The intensity values of the images represent the reconstruction
error in the log scale. Each image from left to right, represents the recovery
error for our measurement scheme as number of chirps $N_c$ is increased and
other measurement schemes as a function of SNR and sparsity level at a fixed
fractional bandwidth
$\frac{\beta}{B} = 0.4$. } \label{fig:RecoveryGuaranteeSNR}
\end{figure*}
Next, we see the influence of noise on sparse target recovery using Basis
pursuit de-noise employing our measurement scheme. We fix fractional bandwidth
$\frac{\beta}{B} = 0.4$ and vary the noise variance as well as the number of
targets in the scene. In Figure~\ref{fig:RecoveryGuaranteeSNR} the intensity of
the image represents the mean square error in dB scale.
Figure~\ref{fig:RecoveryGuaranteeSNR} shows that as the number of chirps
increase, the performance achieved by our scheme in terms of the mean square
error approaches the mean square error achieved by the random Gaussian matrix
$\mbf{G}$. The reconstruction error for Toeplitz matrix $\mbf{T} $ is
marginally better at lower values of $SNR \approx 10 dB$ compared to our
measurement scheme. 

\section{Proofs}\label{sec:Proofs} 
\begin{IEEEproof}
	[Proof of lemma~\ref{lem:opNormTailBound} ] The result can be obtained
by direct application of lemma~\ref{lem:MatBernstein}, using the value of $L =
\sqrt{\frac{N}{M N_c}}$ and $\nu(\mbf{A}) = \frac{N}{M}$ obtained from
lemma~\ref{lem:normBound} and lemma~\ref{lem:varianceBound}, respectively. The
upper bound on the tail probability is given as 
	\begin{align*}
		P\pb{\norm{\mbf{A}}_{op} \geq t} \leq \pb{N + M} \exp
\pb{\frac{-t^2/2}{\sqrt{\frac{N}{M N_c}} \frac{t}{3} + \frac{N}{M} }}. 
	\end{align*}
	By plugging in $t=2\sqrt{\frac{N \log \pb{N +M }}{M}}$, we
get the result in \eqref{eq:MeasMatOpNorm}. For the tail probability to decay,
we require that $\alpha_2 > 1$. This gives us the condition that $1/3
\sqrt{\frac{\log\pb{N +M}}{N_c}} <0.5$, which implies $N_c \geq \frac{4}{9}
\log\pb{N +M}$. Similarly, using the estimates $L$, and $\nu(\mbf{A})$ we can
bound the expected value of the operator norm of $\mbf{A}$ as given in
lemma~\ref{lem:MatBernstein}. 
\end{IEEEproof}
\begin{IEEEproof}
	[Proof of lemma~\ref{lem:minColumnNormTailBound}] The norm of a column m
of the sensing matrix $\mbf{A}$ can be written as follows 
	\begin{align*}
		\norm{\mbf{A}(m)}_2^2 &= \mbf{c}^{*} \mbf{B} \mbf{c}, 
	\end{align*}
	where $\mbf{B} = \mbf{G}_m^{*} \mbf{F}^{*} \mbf{E}_m^{*} \mbf{E}_m
\mbf{F} \mbf{G}_m$, and $\mbf{c} \in \mbb{R}^N$ is a sequence of random
variables that selects and scales a subset of the chirp waveforms. It can be
verified that the diagonal elements of matrix B are as follows 
	\begin{align*}
		B_{i,i} = \frac{1}{N_c }, i=1,\cdots,N. 
	\end{align*}
	Since the random variables $c_i$ are independent and $\Expec{c_i} = 0$,
the off-diagonal terms vanish and we get 
	\begin{align*}
		\Expec{\norm{\mbf{A}(m)}_2^2} &= \sum_{i=1}^{N}
\Expec{\abs{c_i}^2} B_{i,i}, \nn \\
		&= \frac{1}{N} \sum_{i=1}^{N} 1,\nn \\
		\Expec{\norm{\mbf{A}(m)}_2^2} &= 1. 
	\end{align*}
	Using results from lemma~\ref{lem:FroNormOpNormMatrix} and
lemma~\ref{lem:HansonWrightCom} along with the result on sub-Gaussian norm from
lemma~\ref{lem:subGaussNormRes}, we have 
	\begin{align}
		\label{eq:normTail} &P\pb{\abs{\norm{\mbf{A}(m)}_2^2- 1} > t}
\leq \nn \\
		& 4 \exp \pb{-\frac{d}{\frac{N_c}{N}^{\pb{\frac{2}{q^{*}}}}
\frac{1}{N_c q^{*}} \ceil{\frac{N}{M}} }\min
\pb{\frac{t^2}{\frac{N_c}{N}^{\pb{\frac{2}{q^{*}}}} \frac{M }{N_c q^{*}}
\ceil{\frac{N}{M}} } ,t } }, \\
		 & P\pb{\norm{\mbf{A}(m)}_2^2 < 1 - t} \leq
\nn\\
		&4 \exp \pb{-\frac{d}{\frac{N_c}{N}^{\pb{\frac{2}{q^{*}}}}
\frac{1}{N_c q^{*}} \ceil{\frac{N}{M}} }\min
\pb{\frac{t^2}{\frac{N_c}{N}^{\pb{\frac{2}{q^{*}}}} \frac{M }{N_c q^{*}}
\ceil{\frac{N}{M}} } ,t } } \nn
	\end{align}
	The concentration inequality for the minimum value of norm of any column
can be written as for any $t \in [0,1] $, 
	\begin{align}
		&P\pb{\min_m \norm{\mbf{A}(m)}_2^2 \geq 1 -t} \geq 1 - N P\pb{
\norm{\mbf{A}(m)}_2^2 \leq 1 - t }, \nn \\
		& P\pb{\min_m \norm{\mbf{A}(m)}_2^2 \leq 1 - t} \leq N P\pb{
\norm{\mbf{A}(m)}_2^2 \leq 1 - t }. \nn 
	\end{align}
	Let $t = \epsilon$ for any $\epsilon \leq
\frac{\pb{\frac{N_c}{N}}^{\frac{2}{q^{*}} -1 }}{q^{*}}  \in (0,1]$.
Using the approximation $\ceil*{\frac{N}{M}} \approx \frac{N}{M}$, we get the
required result. 
\end{IEEEproof}
\begin{IEEEproof}
	[Proof of lemma~\ref{lem:mutualCoherenceTailBound}] We can express the
inner-product between any two columns $m_1$ and $m_2$ of sensing matrix as 
	\begin{align*}
		\inProd{\mbf{A}(m_1)}{\mbf{A}(m_2)} = \mbf{c}^T \bar{\mbf{B}}
\mbf{c}, 
	\end{align*}
	where $\bar{\mbf{B}}=\mbf{G}_{m_1}^{*} \mbf{F}^{*} \mbf{E}_{m_1}^{*}
\mbf{E}_{m_2} \mbf{F} \mbf{G}_{m_2}$. The diagonal terms of the matrix
$\bar{\mbf{B}}$ is given as 
	\begin{align*}
		&\bar{B}_{i,i} = D_M\pb{\frac{m_1- m_2}{N}} \frac{1}{N_c }
\exp\pb{\frac{2\pi j ( m_1-m_2)(i-1)}{N}}, \\
		&D_M\pb{\frac{m_1- m_2}{N}} =\frac{1}{M}\sum_{t=0}^{M-1}
\exp\pb{2\pi j \pb{\frac{m_1 - m_2}{N}}t } 
	\end{align*}
	where $i=1,\cdots,N$. By using the fact that $c_i$ are zero mean
independent random variables, we obtain the following expression for
$\Expec{\mbf{c}^T \bar{\mbf{B}} \mbf{c} }$ 
	\begin{align*}
		&\Expec{\mbf{c}^T \bar{\mbf{B}} \mbf{c}} = \sum_i \Expec{c_i^2}
\bar{B}_{i,i} \nn \\
		&= D_M\pb{\frac{m_1- m_2}{N}}\sum_{i=1}^N\exp\pb{\frac{2\pi j (
m_1-m_2)(i-1)}{N}} \nn \\
		&=0. 
	\end{align*}
	Using the results from lemma~\ref{lem:FroNormOpNormMatrix} and
lemma~\ref{lem:HansonWrightCom} along with the sub-Gaussian norm result from
lemma~\ref{lem:subGaussNormRes}, and making the approximation
$\ceil*{\frac{N}{M}} \approx \frac{N}{M}$, we see that there $\exists \ \alpha_3
>0$ such that 
	\begin{align*}
		&P\pb{\abs{\inProd{\mbf{A}(m_1)}{\mbf{A}(m_2)}} >
\alpha_3 \sqrt{\frac{\log N}{M}}} \leq \nn\\
		& 4\exp \pb{\frac{-q^{*}d\alpha_3 \log
N}{\frac{N_c}{N}^{\pb{\frac{2}{q^{*}}-1 }} }h(N) }, 
	\end{align*}
	where $h(N) = \min \pb{\frac{q^{*}\alpha_3
}{\frac{N_c}{N}^{\pb{\frac{2}{q^{*}} -1}} } ,\sqrt{\frac{M}{\log N}} }$. Using
the fact that $M \geq (\log N)^3$ and $\ceil*{\frac{N}{M}} \approx
\frac{N}{M}$, we get 
	\begin{align}
		\label{eq:inPordTailBound}
&P\pb{\abs{\inProd{\mbf{A}(m_1)}{\mbf{A}(m_2)}} > \alpha_3
\sqrt{\frac{\log N}{M}}} \nn \\
		& \leq 
		\begin{cases}
			\frac{4}{N^{ u_1}} &\mbox{ if }\log N >
\frac{q^{*}}{(\frac{N_c}{N})^{(2/q^{*}-1)}} \alpha_3 \\
			\frac{4}{N^{ u_2}} & \mbox{otherwise, } 
		\end{cases}
	\end{align}
	where 
	\begin{align*}
		&u_1 = d \pb{\frac{q^{*}\alpha_3
}{\frac{N_c}{N}^{\pb{\frac{2}{q^{*}}-1}}}}^2,\\
		&u_2 = \frac{q^{*}\alpha_3
}{\frac{N_c}{N}^{\pb{\frac{2}{q^{*}}-1}}}d \log N. 
	\end{align*}
	The concentration inequality for the coherence of matrix $\mbf{A}$ can
be obtained by using the following inequality 
	\begin{align*}
		\mu(\mbf{A}) &= \max_{m_1, m_2}
\frac{\abs{\inProd{\mbf{A}(m_1)}{\mbf{A}(m_2)}}}{\norm{\mbf{A}(m_1)}\norm{\mbf{A
}(m_2)} } \\
		&\leq \max_{m_1, m_2} \abs{\inProd{\mbf{A}(m_1)}{\mbf{A}(m_2)}}
\max_m \frac{1}{\norm{\mbf{A}(m)}^2 } 
	\end{align*}
	\begin{align}
		&P\pb{\mu\pb{\mbf{A}} \geq  \frac{\alpha_3}{1 - \epsilon}
\sqrt{\frac{\log\pb{N}}{M}} } \nn \\
		&\leq P\pb{ \max_{m_1,m_2
}\abs{\inProd{\mbf{A}(m_1)}{\mbf{A}(m_2)}} \geq
\alpha_3\sqrt{\frac{\log\pb{N}}{M}} } \nn \\
		& + P\pb{\min_m \norm{\mbf{A}(m)}^2_2 \leq  1 -  \epsilon }
\nn \\
		&\leq \frac{N^2}{2}P\pb{
\abs{\inProd{\mbf{A}(m_1)}{\mbf{A}(m_2)}} \geq
\alpha_3\sqrt{\frac{\log\pb{N}}{M}} } \nn \\
		& + P\pb{\min_m \norm{\mbf{A}(m)}^2_2 \leq 1 - \epsilon }
\nn 
	\end{align}
	Using \eqref{eq:normColumnTail} and \eqref{eq:inPordTailBound} in the
above expression, we get the result in \eqref{eq:mutualCohTail}. 
\end{IEEEproof}
\begin{IEEEproof}
	[Proof of Theorem~\ref{thm:MultiChirpsRecGuarantee}] Using $M \geq
\log(N)^3$ and $\log N > \frac{q^{*}}{(\frac{N_c}{N})^{(2/q^{*}-1)}} \alpha_3$
in \eqref{eq:mutualCohTail}, the coherence condition given in
\cite{candesPlanL1} is satisfied with high probability as shown below 
	\begin{align}
		\label{eq:cohBoundFinal} \mu\pb{\mbf{A}} &= \mc{O}
\pb{\frac{1}{\log N}} \nn\\
		\text{ w.p. } & p_1 \geq 1 -\frac{2}{N^{u_1-2 }} + 4N
\exp\pb{-d M\bar{\epsilon}^2 }, 
	\end{align}
	where $\alpha_3 >0$ is a constant independent of N and M, $\epsilon >0,
> 1,\epsilon_1 \in (0,1) $,
	\[u_1 = d \pb{\frac{q^{*}\alpha_3
}{\frac{N_c}{N}^{\pb{\frac{2}{q^{*}}-1}}}}^2,\]
	and $q^{*} = \max\pb{1,2\log\pb{\frac{N}{N_c}}}$. This establishes the
condition in \eqref{eq:numMeasurementCondition}. We also note that the exponent
in the probability tail bound in \eqref{eq:numMeasurementCondition} depends on
$\frac{N_c}{N}$ as the function $\frac{N_c}{N}^{\pb{\frac{2}{q^{*}}-1}}$ is an
increasing in $N_c \in [0,1/exp(1)]$ but decreases in $N_c \in [1/exp(1),1]$. In
addition to this, we also believe that $\alpha_3$ decreases as $N_c$ increases
linearly with $N$, which is also verified numerically in
section~\ref{sec:Results}. This leads to part of the condition in
\eqref{eq:numChirpsCondition} given as $\frac{N_c}{N} \geq \nu \ll 1$. \par The
measurement matrix in our analysis does not have unit norm columns and in order
to apply lemma~\ref{lem:recoveryGuaranteePlan}, we normalize the columns. We
follow the approach similar to \cite{SparseMIMORadar}. Let $\mbf{D} \in
\mbb{R}^{N \times N}$ diagonal matrix with diagonal entries corresponding to the
norm of the column of $\mbf{A}$ given by 
	\begin{equation*}
		D_{i,i} = \norm{\mbf{A}(i)}_2. 
	\end{equation*}
	The measurement model can be modified as 
	\begin{align*}
		\mbf{y} = \hat{\mbf{A}} \mbf{z} + \mbf{w}, 
	\end{align*}
	where $\hat{\mbf{A}} = \mbf{A}\mbf{D}^{-1}$ and $\mbf{z} =
\mbf{D}\mbf{x}$. Next, we obtain the probability tail bound for the operator
norm of the measurement matrix $\hat{\mbf{A}}$ with $\ell_2$ normalized columns.
Using lemma~\ref{lem:operatorNormBounds}, we have $\forall \epsilon >0,
\epsilon \in (0,1)$, independent of N and M, 
	\begin{align}
		&P\pb{ \norm{\hat{\mbf{A}}}_{op} \geq \frac{2}{\sqrt{1
-\epsilon}} \sqrt{\frac{N \log \pb{N + M}}{M}}} \nn
\\
		& \leq P\pb{\norm{\mbf{A}}_{op} \geq 2
 \sqrt{\frac{N \log \pb{N + M}}{M}}} \nn \\
		& + P\pb{\min_m D_{m,m} \leq \sqrt{1- \epsilon}} \nn\\
		& \leq \pb{\frac{1}{N +M} }^{\alpha_1 -1}+  4N
\exp\pb{-d M\bar{\epsilon}^2 }, \nn 
	\end{align}
	where 
	\begin{align*}
		&\alpha_1 = \frac{2 }{1+\frac{2}{3}
\sqrt{\frac{\log \pb{N + M}}{N_c}}}, \\
		&N_c \geq \frac{4}{9} \log(N + M), \\
		&\bar{\epsilon} =  \pb{\epsilon
\frac{q^{*}}{\pb{\frac{N_c}{N}}^{\frac{2}{q^{*}} -1 }} }. 
	\end{align*}
	
	Therefore, 
	\begin{align}
		\label{eq:OpNormBoundFinal} \norm{\hat{\mbf{A}}}_{op} &\leq
\frac{2}{\sqrt{1-\epsilon}} \sqrt{\frac{N \log \pb{N + M}}{M}}
\\
		\text{ w.p. } & p_2 \geq 1 - \pb{\frac{1}{N +M} }^{\alpha_1 -1}+
4N \exp\pb{-d M\bar{\epsilon}^2 }. \nn 
	\end{align}
	This gives us the condition in \eqref{eq:numChirpsCondition}. Using
\eqref{eq:OpNormBoundFinal} in \eqref{eq:sparsityBoundPlan} we obtain the
condition in \eqref{eq:sparsityCondition}. \par Next, we establish that the
measurement matrix does not reduce the absolute value of non-zero entries of the
sparse vector x below the noise level. 
	\begin{align}
		&P\pb{\min_i D_{i,i}\abs{x_i} \leq 8\sigma\sqrt{2\log N} } \leq
N P\pb{D_{i,i} \leq \sqrt{1 - \epsilon} } \nn \\
		& \leq 4N \exp\pb{-d M\bar{\epsilon}^2 }. \nn 
	\end{align}
	Therefore, we have 
	\begin{align}
		\label{eq:VectorNormBound} &\min_i \abs{z_i} \geq 8\sigma
\sqrt{2 \log N} \\
		\text{w.p. } &p_3 \geq 1- 4N \exp\pb{-d M\bar{\epsilon}^2 }.
\nn 
	\end{align}
	We define the following events associated with a realization of
measurement matrix $\mbf{A}$ 
	\begin{align*}
		&\Xi_1 : \mu\pb{\mbf{A}} = \mc{O} \pb{\frac{1}{\log N}} \\
		&\Xi_2 : \norm{\hat{\mbf{A}}}^2_{op} \leq \frac{c_0
N}{K_{\max}\log N } \\
		&\Xi_3 : \min_i \abs{z_i} \geq 8\sigma \sqrt{2 \log N}. \\
		&\Xi_4 : \mbox{successful support recovery for a fixed sensing
matrix}. 
	\end{align*}
	Let $\Xi$ be the event that the sampled measurement matrix satisfies the
conditions required for successful recovery and recovers a K-sparse vector
$\mbf{x}$ selected from the target model. This implies 
	\begin{align}
		\label{eq:finalUBound}P\pb{\Xi} &\geq P\pb{\Xi_4 \g \Xi_1 \cap
\Xi_2 \cap \Xi_3}\times \nn \\
		& (1-P\pb{\Xi^c_1} - P\pb{\Xi^c_2} - P\pb{\Xi^c_3}). 
	\end{align}
	Using result from Lemma~\ref{lem:recoveryGuaranteePlan} for $P\pb{\Xi_4
\g \Xi_1 \cap \Xi_2 \cap \Xi_3}$,\eqref{eq:OpNormBoundFinal},
\eqref{eq:VectorNormBound} and \eqref{eq:cohBoundFinal} in
\eqref{eq:finalUBound}, we get the desired recovery guarantee. 
\end{IEEEproof}
\begin{IEEEproof}[Proof of Theorem~\ref{thm:RIPCondition}] 
 Using $\delta_K \leq \delta
+\epsilon$ in \eqref{eq:RIPCondition}, it can be deduced that the eigen values
of $\mbf{A}^{*}_{\Gamma}\mbf{A}_{\Gamma},\forall \Gamma$ are $\in \sqb{1 -
\delta - \epsilon, 1 + \delta + \epsilon}$ such that $card\pb{\Gamma}
\leq K$. This can be translated into a condition on the elements of the Gramian
matrix $\mbf{A}^{*} \mbf{A}$ using lemma~\ref{lem:eigValBound} given as the
following events $\Xi_5: \bigcap_{\st{m_1,m_2}{m_1\neq m_2}}
\cb{\abs{\inProd{\mbf{A(m_1)}}{\mbf{A(m_2)}}} \leq \frac{\delta}{K} }$ $
\Xi_6: \bigcap_{m=1}^M \cb{\abs{\norm{\mbf{A}(m)}_2^2  - 1} \leq \epsilon }.$

Let $\Xi$ be the event denoting that the RIP condition of order $K$ is
satisfied. Then we have
\begin{align}
 \label{eq:RIPBound}P(\Xi^c) & \leq 1 - \frac{N^2}{2}
P\pb{\abs{\inProd{\mbf{A(m_1)}}{\mbf{A(m_2)}}} \geq \frac{\delta}{K} }  \nn \\
& \qquad - N P\pb{\abs{\norm{\mbf{A}(m)}_2^2  - 1} \geq \epsilon}
\end{align}
Using the result from \eqref{eq:normTail} and using the fact that
we can choose $N_c$ such that $\epsilon \leq \frac{q^{*}
}{\frac{N_c}{N}^{\pb{\frac{2}{q^{*}}-1}}}$, we get
\begin{align} \label{eq:normEq1}
 P\pb{\abs{\norm{\mbf{A}(m)}_2^2  - 1} \geq \epsilon} \leq 4 \exp\pb{-d
\pb{\epsilon\frac{q^{*}
}{\frac{N_c}{N}^{\pb{\frac{2}{q^{*}}-1}}}}^2 M } 
\end{align}
Similarly, using results from lemma~\ref{lem:mutualCoherenceTailBound} we can
obtain
\begin{align} \label{eq:inProdEq1}
 & P\pb{\abs{\inProd{\mbf{A(m_1)}}{\mbf{A(m_2)}}} \geq 
\frac{\delta}{K} } \leq \nn \\
&\qquad 4 \exp\pb{-d M\frac{\delta^2}{K^2}\pb{\frac{q^{*}
}{\frac{N_c}{N}^{\pb{\frac{2}{q^{*}}-1}}}}^2   }. 
\end{align}
Using \eqref{eq:normEq1}, \eqref{eq:inProdEq1} and the condition that $M \geq a
\delta^{-2} K^2 \log N$ in \eqref{eq:RIPBound} we get the desired results.
\end{IEEEproof}

\section{Conclusion}\label{sec:Conclusion} In this work we have shown that
structured compressed sensing matrices with random components can be realized 
in radar (delay estimation) setting using an  LFM waveform modulated by a random
sparse 
multi-tone signal in transmit and a simple traditional analog receiver
structure. 
We provide recovery guarantees for the proposed compressive sensor comparable
to 
random Toeplitz/Circular matrices with much larger number of random elements. 
The proposed scheme is well matched to practical implementation utilizing small
number of random 
parameters and uniform sampling ADCs on receive.

 A potential direction for  future research is  to investigate the effectiveness
of multi-chirp waveforms in the
multiple output multiple input (MIMO) setting with multiple transmit and receive
antenna elements 
for estimating support of targets in angle and range
domain~\cite{SparseMIMORadar}.
We note that in this setting each transmitter can use a single  chirp with a
random frequency offset and superposition is achieved at each receiver as the
waveforms reflected from the scene is naturally summed at each receiver. 

\appendix[Some Useful Lemmas] 
\begin{lemma}\label{lem:eigValBound}
Given a complex matrix $M \in \mbb{C}^{n \times n}$ with $C_i = M_{i,i}$, and
$R_i = \sum_{j \neq i} \abs{M_{i,j}}$ then we have
\begin{align}
	\lambda_i \in \cup_{i=1}^n D(C_i,R_i), \forall i=1,\cdots,n,
\end{align}
where $D(c,r)$ is a disc with center c and radius r, and $\lambda_i$ are the
eigen-values of M.
\end{lemma}
We restate the theorem given in
\cite{SparseMIMORadar}, which was extended to the complex setting in
\cite{candesPlanL1}, that give the conditions on the measurement matrix for
successful recovery when using $\ell_1$ penalized optimization methods. 
\begin{lemma}
	\label{lem:recoveryGuaranteePlan} For a measurement model $\mbf{y}=
\mbf{A}\mbf{x} +\mbf{w}$, where $\mbf{A} \in \mbb{C}^{M \times N}$ has unit
$\ell_2$ norm columns, $\mbf{x}$ is drawn from a K-sparse model in complex
domain and $w_i \sim \mc{CN}(0,\sigma^2)$, if the following conditions are
satisfied 
	\begin{align}
		\mu(\mbf{A}) \leq \frac{\alpha_0}{ \log N}, 
	\end{align}
	where $\alpha_0 >0$ is a constant independent of the dimensions of the
problem; also, if 
	\begin{align}
		\label{eq:sparsityBoundPlan} K \leq K_{\max}\pb{\mbf{A}} =
\frac{\alpha_1 N}{\norm{\mbf{A}}_{op}^2 \log\pb{N} }, 
	\end{align}
	for some $\alpha_1 > 0$, and 
	\begin{align}
		\min_{k \in I} \abs{x_k} > 8\sigma\sqrt{2 \log\pb{N}}, 
	\end{align}
	then the solution $\hat{\mbf{x}}$ of \eqref{eq:l1Relaxation} has the
same support as the unknown sparse vector $\mbf{x}$ and relative error is
bounded as shown below 
	\begin{align}
		&\emph{supp}(\hat{\mbf{x}}) = \emph{supp}(\mbf{x}) \\
&Pr\pb{\frac{\norm{\hat{\mbf{x}}-\mbf{x}}_2}{\norm{\mbf{x}}_2}\leq
\frac{\sigma\sqrt{3N}}{\norm{\mbf{y}}_2}} \nn \\
		&\geq 1- 2N^{-1}(2\pi \log \pb{N} + KN^{-1}) - \mc{O}
\pb{N^{-2\log 2}}. 
	\end{align}
\end{lemma}
\begin{lemma}
	[Matrix Bernstein inequality \cite{MatrixConcentration}]
\label{lem:MatBernstein} Let $\mbf{A}_i$ be a sequence of i.i.d. random
matrices. For a random matrix expressed as $\mbf{A} = \sum_i \mbf{A}_i$ we have 
	\begin{align}
		&P\pb{\norm{\mbf{A}}_{op} \geq t} \leq \pb{d_1 + d_2} \exp
\pb{\frac{-t^2/2}{\frac{Lt}{3} + \nu\pb{\mbf{A}} }}, \\
		&\norm{\mbf{A}_i}_{op} \leq L, \forall i=1,\cdots, D \nn \\
		&\nu\pb{\mbf{A}} = \max
\pb{\Expec{\mbf{A}\mbf{A}^{*}},\Expec{\mbf{A}^{*}\mbf {A}} }, \nn 
	\end{align}
	where $\mbf{A}_i \in \mbb{C}^{d_1 \times d_2}$. The expected value of
the operator norm of $\mbf{A}$ is bounded by 
	\begin{align}
		\Expec{\norm{\mbf{A}}_{op} }\leq \sqrt{2 \nu\pb{\mbf{A}}
\log\pb{d_1 +d_2}} + \frac{L \log\pb{d_1 +d_2}}{3}. 
	\end{align}
\end{lemma}
\begin{lemma}
	\label{lem:operatorNormBounds} Given a matrix $\mbf{A}$, the operator
norm of the matrix $\hat{\mbf{A}} = \mbf{A}\mbf{D}^{-1}$ is bounded by the
following inequality 
	\begin{align}
		\frac{\norm{\mbf{A}}_{op} }{\min_i D_{i,i}} \geq
\norm{\hat{\mbf{A}}}_{op} \geq \frac{\norm{\mbf{A}}_{op}}{\max_i D_{i,i}}, 
	\end{align}
	where $\mbf{D}$ is a diagonal matrix with positive diagonal elements. 
\end{lemma}
\begin{IEEEproof}
	For any vector $\mbf{v}$ such that $\norm{\mbf{v}}_2 =1$, we have 
	\begin{align}
		&\bar{\mbf{A}}\mbf{v} =
\frac{\mbf{A}\mbf{D}^{-1}\mbf{v}}{\norm{\mbf{D}^{-1}\mbf{ v}}_2}
\norm{\mbf{D}^{-1}\mbf{v}}_2 = \mbf{A} \mbf{u(v)} \norm{\mbf{D}^{-1}\mbf{v}}_2
\nn \\
		\implies &\norm{\bar{\mbf{A}}\mbf{v}}_2 =
\norm{\mbf{A}\mbf{u(v)}}_2 \norm{\mbf{D}^{-1}\mbf{v}}_2,\nn 
	\end{align}
	where $\mbf{u(v)} =
\frac{\mbf{D}^{-1}\mbf{v}}{\norm{\mbf{D}^{-1}\mbf{v}}_2}$. Since
$\norm{\mbf{v}}_2=1$, we bound the Euclidean norm of $\mbf{D}^{-1}v$ as follows 
	\begin{align}
		\frac{1}{\min_i D_{i,i}} \geq \norm{\mbf{D}^{-1}\mbf{v}}_2 \geq
\frac{1}{\max_i D_{i,i}}.\nn 
	\end{align}
	Using this we get, 
	\begin{align}
		\frac{\norm{\mbf{A}\mbf{u(v)}}_2}{\min_i D_{i,i}} \geq
\norm{\bar{\mbf{A}} \mbf{v}}_2 \geq \frac{\norm{\mbf{A} \mbf{u(v)}}_2}{\max_i
D_{i,i}}.\nn 
	\end{align}
	By taking supremum over $\mbf{v}$ in the space of unit norm vectors we
obtain the desired result. 
\end{IEEEproof}
\begin{lemma}
	\label{lem:normBound} For the matrix $c_i \mbf{H}_i \bar{\mbf{A}}
\mbf{D}_i \in \mbb{C}^{M\times N}$ in \eqref{eq:individualChirpMatrices}, we
have 
	\begin{align}
		\label{eq:chirpMatOpNorm} \norm{c_i \mbf{H}_i \bar{\mbf{A}}
\mbf{D}_i }_{op} \leq \sqrt{\frac{N}{M N_c} }, 
	\end{align}
	$\forall i=1,\cdots, N$ 
\end{lemma}
\begin{IEEEproof}
	By sub-multiplicativity property of the operator norm we have, 
	\begin{equation*}
		\norm{c_i \mbf{H}_i \bar{\mbf{A}} \mbf{D}_i }_{op} \leq
\abs{c_i} \norm{ \mbf{H}_i}_{op} \norm{\bar{\mbf{A}}}_{op} \norm{ \mbf{D}_i
}_{op}. 
	\end{equation*}
	Since $\mbf{H}_i$ and $\mbf{D}_i$ are diagonal matrices with complex
exponential entries, it can be shown that $\norm{ \mbf{H}_i}_{op} = \norm{
\mbf{D}_i}_{op} = 1$. Also, we assume that $c_i$ are sub-Gaussian random
variables with $\abs{c_i} \leq 1$. In order to find the operator norm of
$\bar{\mbf{A}}$, we define $\mbf{G}= \bar{\mbf{A}} \bar{\mbf{A}}^{*} \in
\mbf{C}^{M \times M }$ since it is full rank and $ \norm{\bar{\mbf{A}}}^2_{op} =
\norm{\mbf{G}}_{op}$. The entries of matrix $\mbf{G}$ are as follows 
	\begin{align}
		&G(k,k) = \frac{N}{M N_c} \nn \\
		&G(k,l) = \frac{N}{M N_c} \frac{1}{N} \sum_{m=0}^{N-1}
\exp\pb{2\pi j\frac{l-k}{N}m} \nn \\
		&= \frac{N}{M N_c} D_{N}\pb{\frac{l-k}{N}} = 0,\nn 
	\end{align}
	$\forall k,l=0,\cdots, M-1$, such that $k \neq l$, and
$D_{N}\pb{\frac{l-k}{N}} = \frac{1}{N}\sum_{m=0}^{N-1} \exp\pb{2\pi
j\frac{l-k}{N}m} $ is the discrete Dirichlet Kernel. The second term is zero
because the discrete Dirichlet kernel is being evaluated at it's zeros, which
are the Fourier frequency bins $(\frac{n}{N}), n \in \mbb{Z}$. Therefore, 
	\begin{equation*}
		\mbf{G} = \frac{N}{M N_c} \mbf{I}. 
	\end{equation*}
	This implies that the $\norm{\mbf{G}}_{op} = \frac{N}{M N_c}$ and leads
to the result in \eqref{eq:chirpMatOpNorm}. 
\end{IEEEproof}
\begin{lemma}
	\label{lem:varianceBound} For the matrices $ \mbf{P}_i = c_i \mbf{H}_i
\bar{\mbf{A}} \mbf{D}_i \in \mbb{C}^{M\times N}$ given by
\eqref{eq:individualChirpMatrices}, we have 
	\begin{align}
		\label{eq:varBound1}\norm{\sum_{i=1}^{N}
\Expec{\mbf{P}_i^{*}\mbf{P}_i}} &\leq \frac{N}{M}, \\
		\label{eq:varBound2} \norm{\sum_{i=1}^{N} \Expec{\mbf{P}_i
\mbf{P}_i^{*}}} &= \frac{N}{M}. 
	\end{align}
\end{lemma}
\begin{IEEEproof}
	First, we compute the norm of $\mbf{P}_i \mbf{P}_i^{*}$ as it is a
full-rank matrix using $\mbf{D}_i \mbf{D}_i^{*} = \mbf{I}$, $\mbf{H}_i
\mbf{H}_i^{*} = \mbf{I}$ and $\bar{\mbf{A}} \bar{\mbf{A}}^{*} = \frac{N}{M N_c
}\mbf{I}$ we have 
	\begin{align}
		\mbf{P}_i\mbf{P}_i^{*} &= c_i c_i^{*} \mbf{H}_i \bar{\mbf{A}}
\mbf{D}_i \mbf{D}_i^{*} \bar{\mbf{A}}^{*}\mbf{H}_i^{*} \nn\\
		&= c_i c_i^{*} \frac{N}{M N_c} \mbf{I}. \nn 
	\end{align}
	Using the probabilistic model for $c_i$ given in
\eqref{eq:BerRademacher}, we get 
	\begin{align*}
		\Expec{c_i c_i^{*}} = \frac{N_c}{N}. 
	\end{align*}
	We have 
	\begin{align*}
		\sum_{i=1}^{N} \Expec{\mbf{P}_i \mbf{P}_i^{*}} &= \frac{N}{M}
\mbf{I} . 
	\end{align*}
	Applying the operator norm yields the result in \eqref{eq:varBound2}.
Similarly, using the sub-additivity of the operator norm and $\norm{\mbf{H}_i
\bar{\mbf{A}} \mbf{D}_i \mbf{D}_i^{*} \bar{\mbf{A}}^{*} \mbf{H}_i^{*}}_{op} =
\norm{\mbf{D}^{*}_i \bar{\mbf{A}}^{*} \mbf{H}^{*}_i \mbf{H}_i \bar{\mbf{A}}
\mbf{D}_i}_{op}$ we get 
	\begin{align}
		\norm{\sum_{i=1}^{N} \Expec{\mbf{P}_i^{*} \mbf{P}_i} }_{op} &=
\norm{\sum_{i=1}^{N} \Expec{c_i^{*}c_i }\mbf{D}^{*}_i \bar{\mbf{A}}^{*}
\mbf{H}^{*}_i \mbf{H}_i \bar{\mbf{A}} D_i}_{op} \nn \\
		&\leq \sum_{i=1}^{N} \abs{ \Expec{c_i c_i^{*}}} \norm{\mbf{H}_i
\bar{\mbf{A}} \mbf{D}_i \mbf{D}_i^{*} \bar{\mbf{A}}^{*} \mbf{H}_i^{*}}_{op} \nn
\\
		&=\frac{N}{M}. \nn 
	\end{align}
\end{IEEEproof}
\begin{lemma}
	\label{lem:FroNormOpNormMatrix} Let $\mbf{B} = \mbf{G}_m^{*} \mbf{F}^{*}
\mbf{E}_m^{*} \mbf{E}_m \mbf{F} \mbf{G}_m$, and $\bar{\mbf{B}}
=\mbf{G}_{m_1}^{*} \mbf{F}^{*} \mbf{E}_{m_1}^{*} \mbf{E}_{m_2} \mbf{F}
\mbf{G}_{m_2}$ then we have 
	\begin{align}
		& \norm{\mbf{B}}_{op} \leq \ceil*{\frac{N}{M}} \frac{1}{N_c} , &
\norm{\mbf{B}}_{F} \leq \ceil*{\frac{N}{M}} \frac{\sqrt{M}}{N_c}, \\
		& \norm{\bar{\mbf{B}}}_{op} \leq \ceil*{\frac{N}{M }}
\frac{1}{N_c}, &\norm{\bar{\mbf{B}}}_F \leq \ceil*{\frac{N}{M }}
\frac{\sqrt{M}}{N_c}, 
	\end{align}
	where $\ceil{x} =z \in \mbb{Z}$ such that $z \geq x,\ \forall x \in
\mbb{R}$. 
\end{lemma}
\begin{IEEEproof}
	We note that $Rank(\mbf{B}) = Rank(\bar{\mbf{B}}) = M$, and we can
obtain a bound on the Frobenius norm of the matrices as shown below 
	\begin{align*}
		\norm{\mbf{B}}_F &\leq \sqrt{M} \norm{\mbf{B}}_{op}, \\
		\norm{\bar{\mbf{B}}}_F &\leq \sqrt{M}
\norm{\bar{\mbf{B}}}_{op}. 
	\end{align*}
	In order to find the bound on the operator norm we see that 
	\begin{align*}
		\norm{\mbf{B}}_{op} &\leq \norm{\mbf{G}_m}_{op}^2
\norm{\mbf{E}_m}_{op}^2 \norm{\mbf{F}}^2, \\
		\norm{\bar{\mbf{B}}}_{op} &\leq
\norm{\mbf{G}_{m_1}}_{op}\norm{\mbf{G}_{m_2}}_{op}
\norm{\mbf{E}_{m_1}}_{op}\norm{\mbf{E}_{m_2}}_{op} \norm{\mbf{F}}^2. 
	\end{align*}
	Since $\mbf{G}_m$ and $\mbf{E}_m$ are diagonal matrices with complex
exponentials along the principal diagonal, it can be shown that
$\norm{\mbf{G}_m}_{op}^2 = \norm{\mbf{E}_m}_{op}^2 =1$. In order to estimate the
$\norm{\mbf{F}}_{op}$, we see that 
	\begin{align*}
		\mbf{F}^{*} \mbf{F} = \frac{1}{N_c}
		\begin{bmatrix}
			\mbf{V}_{1,1} &\cdots &\mbf{V}_{1,\ceil*{\frac{N}{M}}}
\\
			\mbf{V}_{2,1} &\cdots &\mbf{V}_{2,\ceil*{\frac{N}{M}}}
\\
			\vdots &\vdots &\vdots \\
			\mbf{V}_{\ceil*{\frac{N}{M}},1} &\cdots
&\mbf{V}_{2,\ceil*{\frac{N}{M}}} 
		\end{bmatrix}
	\end{align*}
	Since $p$ is co-prime with $M$, we observe that the N possible frequency
tones circularly get mapped onto M possible aliased sinusoids. Therefore, if
neither $i,j \neq \ceil*{\frac{N}{M}}$, 
	\begin{equation*}
		\mbf{V}_{i,j} = \mbf{I}. 
	\end{equation*}
	If either $i = \ceil*{\frac{N}{M}}$ or $j=\ceil*{\frac{N}{M}}$, then
$\mbf{V}_{i,j} \in \mbb{C}^{M \times M}$ is a partial identity matrix. It can
easily be verified that $\norm{\mbf{F}}_{op}^2 =\frac{1}{N_c} \ceil*{\frac{N}{M
}} $. Using this result along-with the estimate on the bound for the Frobenius
norm, we get the desired results. 
\end{IEEEproof}
\begin{lemma}
	\label{lem:subGaussNormRes} For the sub-Gaussian random variables
described in \eqref{eq:BerRademacher} and \eqref{eq:BerUnifPhase}, the
Sub-Gaussian norm \cite{RandMatricesNonAsymptotics} is as follows 
	\begin{align}
		&\norm{C_i}_{\Psi_2} = \sup_{q \geq 1} \Expec{ \abs{C_i}^{q}
}^{\frac{1}{q}} \frac{1}{\sqrt{q}} \nn \\
		&\norm{C_i}_{\Psi_2} = \pb{\frac{N_c}{N}}^{\frac{1}{q^{*}}
}\frac{1}{\sqrt{q^{*}}}, \\
		&q^{*} = \max\pb{1,2\log\pb{\frac{N}{N_c}}}. \nn 
	\end{align}
\end{lemma}
\begin{IEEEproof}
	For the probability models given in \eqref{eq:BerRademacher} and
\eqref{eq:BerUnifPhase}, we have 
	\begin{align*}
		&\abs{C_i} = 
		\begin{cases}
			0 &\mbox{ with probability } 1 -\frac{N_c}{N} \\
			1 & \mbox{w.p. } \frac{N_c}{N}. 
		\end{cases}
		\nn\\
		&\Expec{\abs{C_i}^{q}}^{\frac{1}{q}} = \pb{
\frac{N_c}{N}}^{\frac{1}{q}}. \nn \\
		\implies &\norm{C_i}_{\Psi_2} = \sup_{q \geq 1}\pb{
\frac{N_c}{N}}^{\frac{1}{q}} \frac{1}{\sqrt{q}}. 
	\end{align*}
	The solution to this optimization problem can be found by taking the
logarithm and solving the unconstrained optimization problem which is given as 
	\begin{align*}
		q^{*} = 2\log\pb{\frac{N}{N_c}}. 
	\end{align*}
	In order to satisfy the constraint, the solution is lower bounded by 1. 
\end{IEEEproof}
\begin{lemma}
	\label{lem:HansonWrightCom} Given a zero mean real random vector
$\mbf{c}$ composed of independent and sub-Gaussian random variables
$c_i,i=1,\cdots,N$ such that $\norm{c_i}_{\Psi_2} \leq K$, $i=1,\cdots,N$, we
have 
	\begin{align}
		\label{eq:HansonWrightComplex} &Pr\pb{\abs{\mbf{c}^{T} \mbf{B}
\mbf{c} - \Expec{\mbf{c}^{T} \mbf{B} \mbf{c}}} > t} \nn\\
		&\leq 4 \exp \pb{-d\min \pb{\frac{t^2}{K^4 \norm{\mbf{B}}_F^2}
,\frac{t}{K^2 \norm{\mbf{B}}_{op}} } }, 
	\end{align}
	where $\mbf{B} \in \mbb{C}^{N \times N}$, $\mbf{c} \in \mbb{R}^{N}$, for
some absolute constant $d >0$ and $\forall t >0$. 
\end{lemma}
\begin{IEEEproof}
	Let $\mbf{B} = \mbf{B}_{R} + i \mbf{B}_{Im}$, where $\mbf{B}_R,
\mbf{B}_{Im} \in \mbb{R}^{N \times N}$. Therefore, using the Hanson-Wright
inequality for real matrices given in \cite{HansonWrightSubGaussian} and the
fact that $\norm{\mbf{B}_R}_{op} \leq \norm{\mbf{B}}_{op}$,
$\norm{\mbf{B}_{Im}}_{op} \leq \norm{\mbf{B}}_{op}$, and $\norm{\mbf{B}_R}_{F}
\leq \norm{\mbf{B}}_{F}$, $\norm{\mbf{B}_{Im}}_{F} \leq \norm{\mbf{B}}_{F}$ we
have 
	\begin{align*}
		Pr&\pb{\abs{\mbf{c}^{T} \mbf{B} \mbf{c} - \Expec{\mbf{c}^{T}
\mbf{B} \mbf{c}}} > t} \leq \nn \\
		&Pr\pb{\abs{\mbf{c}^{T} \mbf{B}_R \mbf{c} - \Expec{\mbf{c}^{T}
\mbf{B}_R \mbf{c}}} > \frac{t}{\sqrt{2}}} + \nn \\
		& Pr\pb{\abs{\mbf{c}^{T}\mbf{B}_{Im} \mbf{c} -
\Expec{\mbf{c}^{T} \mbf{B}_{Im} \mbf{c}}} > \frac{t}{\sqrt{2}}}. 
	\end{align*}
	This gives us the inequality in \eqref{eq:HansonWrightComplex}. 
\end{IEEEproof}

\section*{Acknowledgment}
This research was partially supported by Army Research Office grant
W911NF-11-1-0391 and NSF Grant IIS-1231577.

\bibliographystyle{IEEEtran} 
\bibliography{IEEEabrv,References}

%
\end{document}